\documentclass{aastex631}

\usepackage{amsmath,amssymb}
\usepackage{graphicx}
\usepackage{booktabs}
\usepackage{multirow}
\usepackage{tikz}
\usetikzlibrary{arrows.meta,positioning}

\newcommand{\candpar}[1]{\par\noindent{\normalsize\sffamily #1}\ \ignorespaces}

\shorttitle{Transit survey of 461 TESS M dwarfs}
\shortauthors{Tschudi}

\begin{document}

\title{A uniform transit survey of 461 ExoFOP M-dwarf TOI hosts: Follow-up prioritization and new transit candidates}

\author[0009-0002-7797-1502]{Yohann Tschudi}
\affiliation{Independent Researcher, 69480 Anse, France}
\email{yohann.tschudi@gmail.com}

\begin{abstract}
M dwarfs are favorable hosts for small transiting planets. A transit-detection and statistical-validation pipeline was previously developed and validated at the photometric noise frontier, on M3--M6 dwarfs observed in few sectors. That pipeline is applied uniformly to the full active ExoFOP M-dwarf TOI population, 461 hosts (M0--M6) each hosting a confirmed planet (CP) or planet candidate (PC), to test its recovery and classification of the cataloged signals and to prioritize the candidates for follow-up. The pipeline chains two complementary blind searches merged per host, a full-baseline transit-least-squares scan and a segmented, semi-coherent search across the gapped multiyear baseline. It adds an event-time-coherence test that rejects signals with incoherent per-transit timing (window-function aliases of stellar variability), all-sector TRICERATOPS false-positive probabilities, and Gaia~DR3 background and aperture-localization analysis. Of 193 in-range CP, 165 (85.5\%) are re-detected, none rejected as a false positive, and 147 (76.2\%) reach a Gaia-verified disposition. Of 286 in-range PC, 225 (78.7\%) are re-detected, of which 11 are flagged as false positives and 87 ranked as planet candidates for follow-up. Beyond the catalog, 13 new transit candidates of $0.7$--$2.0\,R_\oplus$ emerge, including two new members of a compact, dynamically stable near-resonant three-planet candidate system; four cataloged single- or dual-transit TOIs have their orbital periods determined, two of them in the optimistic habitable zone. The recoveries and new candidates are released as a uniform, prioritized follow-up target list.
\end{abstract}


\section{Introduction}
\label{sec:intro}

The Transiting Exoplanet Survey Satellite \citep[TESS;][]{Ricker2015} has transformed the census of planets around M dwarfs, the most abundant stellar population in the solar neighborhood \citep{Henry2006}. Transit detection on these faint, active targets remains challenging: stellar variability amplitudes often exceed transit depths \citep{Yaptangco2025}, and the sector-based observing strategy produces gapped light curves that limit sensitivity to long-period planets \citep{Sullivan2015}. Several pipelines address these challenges, including the Science Processing Operations Center \citep[SPOC;][]{Jenkins2016} pipeline, the Quick-Look Pipeline \citep[QLP;][]{Huang2020a, Huang2020b}, and the threshold-based LEO-Vetter \citep{Kunimoto2025}, targeting the detection or vetting steps individually.

\citet{Tschudi2026} developed an automated transit-detection and statistical-validation pipeline and validated it at the photometric noise frontier, on 121 newly enabled M3--M6 dwarfs, where the limited sector coverage makes the search hardest. The present work applies the complete pipeline, unchanged in its detection core, with the vetting adapted to the heterogeneous ExoFOP population (Sect.~\ref{sec:vetting}), uniformly across the full active population of M-dwarf TESS objects of interest \citep[TOIs;][]{Guerrero2021} in the Exoplanet Follow-up Observing Program (ExoFOP), a sample of signals around stars already known to host a planet or candidate, generally observed over more sectors than the noise-frontier sample. Confirmed systems are productive targets for this purpose, since the host is known to harbor at least one planet, its stellar parameters are characterized \citep{Muirhead2018}, and radial-velocity or other follow-up data are often already in hand. The first aim is to establish, on the confirmed planets, that the pipeline recovers and correctly classifies known signals, a consistency test of the full chain; the second is to apply it to the unconfirmed candidates, removing false positives and prioritizing the survivors for follow-up.

A segmented, semi-coherent second search, merged per host with the full-baseline scan, recovers phase-coherent transits that a single full-baseline scan dilutes across the gapped multiyear baseline. A second addition targets a failure mode of the gapped multiyear baseline, where the observing window can stack quasi-periodic stellar variability into spurious transit trains that pass standard statistical validation \citep{Dawson2010}. An event-time-coherence test rejects such signals by requiring that their individual transit times form a coherent clock across the baseline, extending the event-level vetting heritage of the Kepler Robovetter \citep{Mullally2016, Thompson2018, Coughlin2016} and recent TESS vetters \citep{Kunimoto2025}; it is calibrated on a window-function false positive established within the program.

The sample comprises 461 active ExoFOP M-dwarf TOI hosts (M0--M6; Sect.~\ref{sec:target_selection}), of which 193 confirmed planets and 286 planet candidates fall within the $0.5$--$100$\,d search range and set the recovery denominators used throughout.

The paper is organized as follows. Section~\ref{sec:sample} describes the sample, the pipeline, the segmented search, and the event-time check. Section~\ref{sec:recovery} presents the confirmed-planet recovery that validates the solution, and Section~\ref{sec:pc_recovery} the candidate recovery with the false positives the checks remove. Section~\ref{sec:candidates_revised} presents the surviving new candidates; Sections~\ref{sec:discussion} and \ref{sec:conclusions} give the limitations and conclusions.

\section{Sample and methods}
\label{sec:sample}

The detection, vetting, and verification pipeline of \citet{Tschudi2026}, described and validated in full there on 121 newly enabled M3--M6 dwarfs (only recently searchable through the accumulation of TESS Cycle~6+ coverage), is applied here across the full M0--M6 ExoFOP M-dwarf population. This section recaps the pipeline (Fig.~\ref{fig:pipeline_flowchart}, Appendix~\ref{app:flowchart}) and sets out the components added for the present survey: a segmented, semi-coherent second search pass and the per-host merge of the two detection engines (Sect.~\ref{sec:segmented}); an event-level coherence screen (Sect.~\ref{sec:screening}); vetting adaptations for the heterogeneous ExoFOP population, including the all-sector median false-positive-probability protocol (Sect.~\ref{sec:vetting}); and a photometric aperture-localization test (Sect.~\ref{sec:verification}). Every operational threshold of these added components was frozen, before any application to the survey, on the ten-system validation set of Sect.~\ref{sec:validation_known} (Table~\ref{tab:frozen_thresholds}), chosen to span the survey's regimes: single- and multiplanet hosts (up to four planets, including a near-commensurable pair), quiet to highly active stars, and 2 to 37 observed sectors.

\subsection{Target Selection}
\label{sec:target_selection}

The sample comprises the 461 M dwarfs in the TESS Input Catalog \citep[TIC;][]{Stassun2019} hosting at least one TOI with an active TESS Follow-up Observing Program Working Group (TFOPWG) disposition, confirmed planet (CP) or planet candidate (PC), in ExoFOP-TESS; the dispositions and all derived counts use the pinned 2026 June 29 snapshot. The sample was searched in two campaigns probing complementary spectral ranges (Table~\ref{tab:survey_overview}). This sample does not overlap the 121 newly enabled M3--M6 dwarfs searched in \citet{Tschudi2026}. The snapshot contained 611 M-dwarf TOI hosts; the 150 whose TOIs carried only known-planet (KP, 23), false-positive or false-alarm (FP/FA, 102), or ambiguous (APC, 25) dispositions were not searched, leaving the active CP/PC population. The recovery of known transiting planets is characterized in detail on the ten benchmark systems of Sect.~\ref{sec:validation_known}, which are themselves members of this sample. No brightness, sector-count, or stellar-quality filter was applied. The M0--M2 campaign contains 269 targets with $T_{\mathrm{eff}} = 3500$--$4000$\,K (76 CP and 195 PC hosts); the M3--M6 campaign contains 192 targets with $T_{\mathrm{eff}} < 3500$\,K (86 CP and 108 PC hosts).

\begin{deluxetable}{lcc}
\tablecaption{Survey overview.\label{tab:survey_overview}}
\tablehead{\colhead{Parameter} & \colhead{M0--M2} & \colhead{M3--M6}}
\startdata
TICs                       & 269           & 192 \\
$T_{\mathrm{eff}}$ (K)     & 3500--4000    & $<$3500 \\
$R_\star$ ($R_\odot$)      & 0.28--0.71    & 0.10--0.55 \\
CP / PC hosts              & 76 / 195      & 86 / 108 \\
Confirmed planets (in range) & 96          & 97 \\
Evaluable PC signals       & 187           & 99 \\
$N_{\rm sectors}$ (range)  & 1--45         & 2--44 \\
$N_{\rm sectors}$ (median) & 5             & 4 \\
$T_{\rm mag}$ range        & 7.5--14.5     & 8.0--14.5 \\
\enddata
\tablecomments{CP = Confirmed Planet; PC = Planet Candidate (ExoFOP-TESS dispositions). CP/PC hosts count unique TICs; their sum exceeds the TIC total because two targets in each campaign host both CP and PC signals. ``Confirmed planets (in range)'' and ``Evaluable PC signals'' count individual signals with a period in the $0.5$--$100$\,d search range (some hosts have multiple planets); these are the recovery denominators. Under the 2026 June 29 disposition snapshot the sample contains 311 PC signals in total, of which 286 ($187 + 99$) are in range; 202 confirmed planets in total, of which 193 ($96 + 97$) are in range. Campaign assignment uses the TIC effective temperature \citep{Stassun2019}; the few hosts with a higher spectroscopic value are assigned by their TIC value.}
\end{deluxetable}

\subsection{Detection, Vetting, and Verification Pipeline}

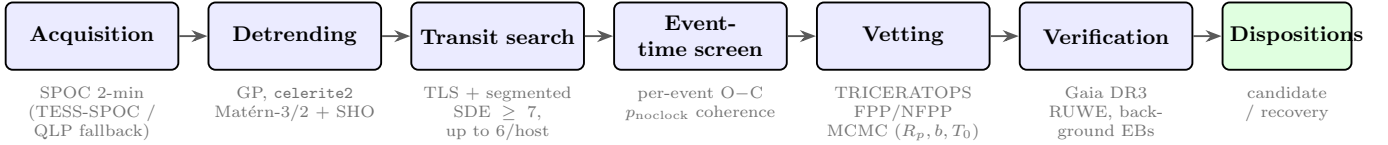
\begin{figure*}[!t]
\centering
\resizebox{\textwidth}{!}{%
\begin{tikzpicture}[
  node distance=0.42cm,
  stage/.style={draw, rounded corners=3pt, minimum height=1.0cm, text width=2.35cm,
                font=\small, align=center, fill=blue!8, thick},
  meth/.style={font=\scriptsize, text=black!55, align=center, text width=2.45cm},
  arr/.style={-{Stealth[length=2.4mm]}, thick, black!65},
]
\node[stage] (acq) {\textbf{Acquisition}};
\node[stage, right=of acq] (gp)  {\textbf{Detrending}};
\node[stage, right=of gp]  (tls) {\textbf{Transit search}};
\node[stage, right=of tls] (scr) {\textbf{Event-time screen}};
\node[stage, right=of scr] (vet) {\textbf{Vetting}};
\node[stage, right=of vet] (ver) {\textbf{Verification}};
\node[meth, below=0.12cm of acq] {SPOC 2-min\\(TESS-SPOC / QLP fallback)};
\node[meth, below=0.12cm of gp]  {GP, \texttt{celerite2}\\Mat\'ern-3/2 $+$ SHO};
\node[meth, below=0.12cm of tls] {TLS $+$ segmented\\$\mathrm{SDE}\geq7$, up to 6/host};
\node[meth, below=0.12cm of scr] {per-event O$-$C\\$p_{\rm noclock}$ coherence};
\node[meth, below=0.12cm of vet] {TRICERATOPS FPP/NFPP\\MCMC ($R_p,b,T_0$)};
\node[meth, below=0.12cm of ver] {Gaia DR3\\RUWE, background EBs};
\foreach \a/\b in {acq/gp, gp/tls, tls/scr, scr/vet, vet/ver} \draw[arr] (\a) -- (\b);
\node[draw, rounded corners=3pt, right=of ver, minimum height=1.0cm, text width=1.7cm,
      font=\small, align=center, fill=green!12, thick] (dis) {\textbf{Dispositions}};
\draw[arr] (ver) -- (dis);
\node[meth, below=0.12cm of dis, text width=1.9cm] {candidate\\/ recovery};
\end{tikzpicture}%
}
\caption{Overview of the transit-detection and classification pipeline, from TESS photometry to a terminal disposition. Each stage is detailed in Sects.~\ref{sec:detection}--\ref{sec:verification}; the complete disposition logic is given in Fig.~\ref{fig:pipeline_flowchart}.}
\label{fig:pipeline_overview}
\end{figure*}

The pipeline proceeds through the stages of Fig.~\ref{fig:pipeline_overview}, detailed in turn below; the full classification logic, with every terminal disposition, is given in Fig.~\ref{fig:pipeline_flowchart} (Appendix~\ref{app:flowchart}).

\subsubsection{Light-curve Preparation and Detection}
\label{sec:preprocessing}
\label{sec:detection}

For each target, all available TESS light curves are retrieved from the Mikulski Archive for Space Telescopes (MAST\footnote{\url{https://mast.stsci.edu}}), prioritizing SPOC 2-minute Pre-search Data Conditioning Simple Aperture Photometry (PDCSAP) flux \citep{Jenkins2016} with a per-sector fallback to TESS-SPOC and QLP full-frame-image products \citep{Huang2020a, Huang2020b}. Detection follows \citet{Tschudi2026}: sectors are detrended with a \texttt{celerite2} Gaussian process (GP; Mat\'ern-3/2 plus simple-harmonic-oscillator (SHO) kernel; \citealt{ForemanMackey2017}), stitched, and searched with transit least squares \citep[TLS;][]{Hippke2019} iteratively (up to six planets per host, each masked before the next, with a subharmonic step for true periods below the grid floor) over $P = 0.5$--$100$\,d above a signal-detection-efficiency (SDE) threshold of $7$, time-correlated noise penalized by $\mathrm{SDE}_{\mathrm{eff}} = \mathrm{SDE}/\sqrt{\beta_{\rm rn}}$ ($\beta_{\rm rn}\in[1,10]$; \citealt{Pont2006}). Each detection passes the multicheck cascade: period validation against the stellar rotation and the $0.5$, $1.0$, and $13.7$\,d instrumental aliases, and odd-even and secondary-eclipse consistency \citep{Coughlin2016, Thompson2018}.

\subsubsection{Segmented Search and Detection Merge}
\label{sec:segmented}

The SDE contrasts a candidate period against the full ensemble of trial periods. On a gapped, multiyear baseline the observing window raises a forest of alias peaks that inflates that ensemble's noise floor, so a planet whose folded transit is highly significant can still vanish into it. On L\,98-59 (27 sectors over 2\,692\,d), planet d reaches $\mathrm{SDE} = 77.8$ in a windowed search around its period but only $3.7$ in the blind full-baseline scan. A second, semi-coherent pass therefore complements the full-baseline search: it searches each observing block on its own and then combines them, so a faint but coherent train stands out block by block instead of competing against the whole window forest at once.

The light curve is cut at gaps longer than ten days into contiguous blocks (span $\geq 20$\,d), each searched blind at full resolution; the surviving peaks (above $\mathrm{SDE} = 3.5$, at most thirty per block) are grouped across blocks by period. A group spanning at least two blocks is kept only when its recurrence is unlikely to be chance, judged against a null built on the star itself: in $N = 1\,000$ trials the block peaks keep their strengths but have their periods reshuffled on each block's own grid \citep{Ofir2014} and are processed exactly as the observation, alias handling and grouping included; the group advances only when the reshuffled peaks assemble one this convincing in fewer than $1\%$ of trials ($p < 0.01$). Because the null is measured per star, it already accounts for the number of periods searched, correcting for the look-elsewhere effect. Each surviving period is then confirmed on the full baseline with a windowed transit search and required to show mutually consistent per-block depths whose inverse-variance combination reaches $5\sigma$; long-period candidates ($P > 12$\,d), where no single statistic separates a faint genuine train from a window-assembled one, must additionally clear a small jury of depth, folded-S/N, and windowed-SDE cuts. All these thresholds were calibrated once on the validation set and held fixed (Table~\ref{tab:frozen_thresholds}). Because the $p<0.01$ gate is applied across the full sample, it still admits of order ten chance groupings, which the windowed confirmation and depth-consistency stages then remove. As a direct test, each benchmark light curve was block-shift scrambled, every block circularly shifted by an independent random offset so that its window function and per-block periodogram power are preserved but the cross-block transit phase is destroyed. None of 42 such realizations produced a confirmed candidate ($0/42$): the blocks still peak at the true period and pass federation, but the windowed confirmation rejects the phase-decorrelated epochs, which fail to stack to $5\sigma$.

The full-baseline and segmented lists are merged per host. A candidate at an exact integer period ratio ($n = 2$--$6$) to a stronger co-host signal is an iterative-masking alias and is absorbed into that signal's family before matching (genuine near-commensurable planets are never exact: L\,98-59\,d and c sit at a period ratio of $2.019$); the remaining candidates agreeing in period, to within $1\%$ (or $6\%$ beyond 12\,d, where block federation smears it), are merged into one entry at the windowed-confirmation ephemeris. A signal found by only one engine is kept unchanged, the two searches being complementary within a host rather than redundant, and every downstream stage treats single-engine and dual-engine candidates identically.

\subsubsection{Harmonic Cleanup and Event-time Screening}
\label{sec:harmonic_cleanup}
\label{sec:screening}

Because the iterative search runs over the residuals, one true period can reappear as a family of aliases ($2P$ or $3P$ harmonics, a fundamental mistaken for a harmonic, or, on active M dwarfs, BY~Draconis dips folded at unrelated periods). Ranking by detection statistic is unsafe, so each family is resolved on physical evidence as in \citet{Tschudi2026}, analyzing a host's candidates jointly: a cross-masking test (a dependent alias loses more than half its depth once the stronger signal is masked), an activity-scaled parsimony cap, and a mutual-Hill stability test \citep[$3.5\,R_{\mathrm{H}}$;][]{Gladman1993}; a period-ratio commensurability is reported but never demotes a candidate, and only the concurrence of these tests marks an alias. The cleanup retains every confirmed planet of the validation set (Sect.~\ref{sec:validation_known}).

A second screen addresses a failure mode specific to the gapped, multiyear TESS baseline: the observing window can stack quasi-periodic stellar or instrumental structure into a transit-like train at an alias period that survives the standard vetting chain. The risk is greatest near the detection floor, where a candidate is a sparse train of a few faint ($2$--$4\sigma$) events across sector blocks separated by year-long gaps, none inspectable on its own, so a blind detection peak's period is not yet a measurement \citep{Dawson2010, Cooke2021}. Using the events' timing alone, and neither the detection statistic nor the transit depth (both unreliable at this S/N; \citealt{Mullally2018}), the screen tests, for every surviving candidate, whether the individual events keep time like a single ephemeris or scatter as loosely as unrelated noise.

Each candidate is tested at its imposed period and epoch, which are not re-fit. Every predicted event is re-timed locally to the offset that maximizes its depth significance, and the resulting timing dispersion $\omega = \mathrm{median}\,|\mathrm{O}-\mathrm{C}|$ (observed minus calculated) measures the clock coherence: small for a genuine ephemeris, and as large as the transit window for a train assembled by the window function. The verdict is a $p$-value read against a no-clock null measured on the candidate's own light curve, a set of $N = 200$ off-signal ephemerides re-timed by the identical procedure,
\begin{equation}
  p_{\rm noclock} = \frac{1}{N}\,\#\{\,\omega_{\rm null}\le \omega_{\rm cand}\,\},
  \label{eq:pnoclock}
\end{equation}
Equation~(\ref{eq:pnoclock}) is the probability that clockless noise on the star times the events at least as coherently as the candidate. A genuine ephemeris drives $p_{\rm noclock}\to 0$; an alias sits near $0.5$. Building the null per candidate makes the verdict insensitive to the cadence heterogeneity across the survey. Candidates with fewer than five timeable events are set aside as low-sensitivity, and those recovered below the $0.5$\,d search floor (Sect.~\ref{sec:segmented}) are exempt, the timing window there exceeding the orbital period.

\label{sec:screening_calibration}%
The thresholds are set on ground truth from the ten benchmark systems. The sixteen validation confirmed-planet signals (Sect.~\ref{sec:validation_known}) all reach $p_{\rm noclock}\le 0.095$ (fourteen at exactly $0$), whereas known window aliases (the spurious ultra-short-period signals the search produces on GJ\,3473) and synthetic clockless trains injected into the survey light curves all return $p_{\rm noclock}\ge 0.44$. In this wide gap the outcome is insensitive to the exact cuts: \textsc{pass} at $p_{\rm noclock}\le 0.05$, \textsc{fail} at $\ge 0.30$, \textsc{flag} between, \textsc{flag\_lowsens} for fewer than five timeable events. On the fail side, the design case is a $21.78$\,d transit-like train toward TOI-521, aliased by the TESS revisit window from the $20.35$\,d period of the star's non-transiting radial-velocity planet \citep{Lacedelli2026}. This certified window-function false positive fails the screen as expected ($p_{\rm noclock}=0.39$). A \textsc{flag} triggers case-by-case adjudication.

\subsubsection{Statistical Vetting}
\label{sec:vetting}

The survivors are vetted with TRICERATOPS \citep{Giacalone2021} over the TRILEGAL field-star population \citep{Girardi2005}, where FPP and NFPP denote the false-positive and nearby-false-positive probabilities: $\mathrm{FPP} < 1.5\%$ with $\mathrm{NFPP} < 0.1\%$ yields \texttt{VALIDATED}, and $\mathrm{FPP} < 50\%$ a \texttt{LIKELY\_PLANET} \citep{Giacalone2021}. The nearby-false-positive coupling of the likely-planet cut is deferred here to the Gaia verification (Sect.~\ref{sec:verification}); the terminal operational states adopted are \texttt{PENDING} (clean centroid, awaiting observational tests) and \texttt{FALSE\_POSITIVE}.

Three adaptations address the heterogeneity of the ExoFOP population (giant planets, eclipsing binaries, crowded fields, 1--45 sectors); the full rules are in \citet{Tschudi2026} and Appendix~\ref{app:flowchart}. (i) The FPP is evaluated jointly over all sectors in one TRICERATOPS calculation, the field-star catalogs merged and their flux ratios averaged across apertures \citep[Sect.~2]{Giacalone2021}, returning a single FPP and NFPP reported as the median of Monte Carlo draws with their 16--84\% interval; a degenerate draw distribution (Table~\ref{tab:frozen_thresholds}) claims no validation (Sect.~\ref{sec:limitations}), with a single-sector fallback for degenerate crowded fields. Resolved Gaia field stars are retained in the scenario set, and the per-candidate reruns of Sect.~\ref{sec:candidates_revised} add archival contrast curves. (ii) A signal is called \texttt{LIKELY\_EB} only on converging evidence: a deep transit ($> 50\,000$\,ppm with $\mathrm{FPP} \geq 50\%$), or a secondary eclipse \citep{Coughlin2016, Thompson2018, Kunimoto2025} corroborated by a second indicator (elevated eclipsing-binary probability, odd-even depth difference, ellipsoidal variation, or non-negligible NFPP); a strong ellipsoidal variation alone also suffices, whereas an isolated red-noise phase-0.5 feature does not reject a genuine transit. (iii) Validation is capped at \texttt{LIKELY\_PLANET} above $R_p = 8\,R_\oplus$, where planet, brown-dwarf, and low-mass-star radii are degenerate \citep{Giacalone2021}; a centroid--NFPP consensus and stellar-variability guard let an overwhelming NFPP reject a signal only where the event-time screen (Sect.~\ref{sec:screening}) has not established a genuine transit clock.

The vetting stage also refines the transit parameters by a Markov chain Monte Carlo (MCMC) fit of the \texttt{batman} model \citep{Kreidberg2015, ForemanMackey2013} for $R_p/R_\star$, $b$, and $T_0$, with quadratic limb darkening fixed from \citet{Claret2017}. Throughout this paper, quoted planet radii combine the fitted $R_p/R_\star$ posterior with the TIC stellar radius \citep{Stassun2019} at a conservative 5\% floor on $\sigma_{R_\star}/R_\star$; insolations follow $S \propto R_\star^2\,T_{\rm eff}^4/a^2$ ($a$ from Kepler's third law and the TIC $M_\star$), their uncertainty dominated by $\sigma_{R_\star}$ and $\sigma_{T_{\rm eff}}$.

\subsubsection{Gaia Verification}
\label{sec:verification}

Candidates passing this stage undergo Gaia~DR3 background source analysis \citep{GaiaDR3}: sources bright enough to reproduce the transit depth are searched for within the photometric aperture, and the host renormalized unit weight error (RUWE) is checked against the 1.4 single-star threshold \citep{Lindegren2021}. A candidate whose aperture retains a resolved neighbor bright enough to host the transit, which Gaia (resolving only beyond the $0.4\arcsec$ limit of \citealt{Fabricius2021}) cannot itself assign, is flagged \texttt{NEEDS\_HR\_IMAGING}: the transit may lie on the primary, and only high-resolution imaging can resolve the host. A host with an elevated RUWE but no resolved neighbor, the signature of a possible unresolved companion below that limit, instead yields the cautionary \texttt{PLANET\_CANDIDATE\_HEB\_CAUTION} disposition (hierarchical eclipsing binary, HEB), retained for imaging follow-up. For hosts observed in at least eight sectors, a photometric localization test additionally correlates the per-sector transit depth with the per-sector SPOC aperture size \citep[automating the by-hand check of][]{Eisner2020}: an on-target transit is aperture-independent, whereas a depth that scales with the aperture (Spearman $\rho \geq 0.5$ at $p < 0.01$, with at least a twofold depth ratio across apertures) indicts flux entering from a resolved neighbor, and the signal is flagged \texttt{NEB\_APERTURE\_SUSPECT}. For \texttt{PENDING} candidates this analysis governs the disposition (a clean field excludes the background eclipsing binary (BEB) at the resolution limit and yields \texttt{PLANET\_CANDIDATE}); for \texttt{LIKELY\_PLANET} candidates it can only upgrade. Candidates without target pixel files (TPFs) on MAST or with failed world coordinate system (WCS) projections cannot be vetted and are excluded from the statistical results as non-analyzable.

\subsection{End-to-end Validation on Known Systems}
\label{sec:validation_known}

Before application to the survey, the full chain, from acquisition to Gaia~DR3 verification, was validated end-to-end on the same ten benchmark systems as \citet{Tschudi2026}: GJ\,357, GJ\,486, GJ\,3473, Gliese\,12, L\,98-59, LP\,791-18, TOI-406, TOI-700, TOI-782, and TOI-6086, spanning M2--M5 ($T_{\rm mag}=8.7$--$13.6$) and one to four planets per host, with sixteen TESS-detectable confirmed planets.

\begin{deluxetable*}{lll}
\tabletypesize{\scriptsize}
\tablecaption{Frozen operational thresholds of the components added for this survey, calibrated once on the ten-system validation set and held fixed thereafter.\label{tab:frozen_thresholds}}
\tablehead{\colhead{Stage} & \colhead{Parameter} & \colhead{Value}}
\startdata
Segmented search  & block gap / min.\ span        & $>10$\,d / $\geq 20$\,d \\
                  & peak floor / peaks per block  & $\mathrm{SDE}=3.5$ / $\leq 30$ \\
                  & federation                    & $\geq 2$ blocks, $p<0.01$ ($N=1\,000$) \\
                  & windowed confirmation         & stacked $\geq 5\sigma$ \\
                  & $P>12$\,d jury                & depth $\geq 4.5\sigma$, $\mathrm{S/N}\geq 4$, $\mathrm{SDE}\geq 12$ \\
Detection merge   & alias absorption              & integer ratio $n\leq 6$ (tol.\ $10^{-3}$) \\
                  & cross-engine match            & $1\%$ ($6\%$ beyond 12\,d) \\
Event-time screen & \textsc{pass}/\textsc{flag}/\textsc{fail} & $p_{\rm noclock}\leq 0.05$ / $0.05$--$0.30$ / $\geq 0.30$ \\
FPP protocol      & Monte Carlo draws             & $20 \times 10^{6}$ \\
                  & degenerate-fold flag          & $\geq 20\%$ saturated or width $>20$\,pts \\
Aperture test     & off-target localization       & $\rho\geq 0.5$ ($p<0.01$), $\geq 2\times$ depth, $\geq 8$ sectors \\
\enddata
\tablecomments{Thresholds inherited unchanged from \citet{Tschudi2026} are not repeated; the table lists only the thresholds added for the present survey. ``width'' is the $16$--$84\%$ inter-draw FPP interval; ``pts'' are percentage points of FPP.}
\end{deluxetable*}

The merged detection stage recovers all sixteen genuine signals at their archive ephemerides, the two engines being individually incomplete in complementary ways: the full-baseline search misses L\,98-59\,d, whose $7.3814$\,d peak is an exact $2{:}1$ alias of planet c (ratio $2.000$) lying $0.93\%$ from the true planet d at $7.4507$\,d and is absorbed at the merge; the segmented search recovers it blind, federated in all eleven blocks at a stacked $33\sigma$; conversely TOI-700\,d and Gliese\,12\,b are found by the full-baseline search alone. Only two of the ten systems (L\,98-59, TOI-700) lie in the multiyear-gapped regime the segmented search addresses, so this validation measures the method's safety rather than its yield, the survey-scale contribution of the second engine being quantified in Sect.~\ref{sec:recovery}. The seven spurious signals the iterative search raises are each removed at the stage designed for it, the merge, harmonic cleanup, event-time screen, TRICERATOPS vetting, and Gaia and aperture test in turn. The joint folds of TOI-700\,e and TOI-406.01 saturate under the multidraw diagnostic and are flagged as degenerate computations rather than read against the signals, the driver for TOI-700\,e being its published transit-timing variations \citep[RMS ${\sim}1$\,h;][]{Gilbert2023}, which the event-time screen independently measures as a median $|O\!-\!C|$ of 58\,min. A further $1.049$\,d L\,98-59 signal that passes the timing screen is localized onto a resolved neighbor by the aperture test ($\mathrm{NFPP} = 41\%$) and flagged \texttt{NEB\_APERTURE\_SUSPECT}. No confirmed planet is lost and no spurious signal is promoted: the genuine endpoint is 13 \texttt{PLANET\_CANDIDATE} and 3 \texttt{NEEDS\_HR\_IMAGING}, and the event-time screen returns 15 \textsc{pass}, 1 \textsc{flag}, and 0 \textsc{fail} on the genuine signals. Per-system results are in Appendix~\ref{app:validation} (Table~\ref{tab:validation}).

\section{Confirmed planet recovery}
\label{sec:recovery}

\subsection{Transit Search Recovery}
\label{sec:recovery_tls}

A signal is considered recovered when TLS detects a period matching the ExoFOP value within $< 1\%$ relative deviation, or at an integer multiple or sub-multiple alias ($N\!\cdot\!P$ or $P/N$, $N \le 6$) matched to high precision ($< 0.1\%$ in the period ratio). A genuine alias of the same transit train reproduces the integer ratio to a few ppm, whereas an unrelated coincidence matches only to ${\sim}0.5\%$.

\begin{figure}[!t]
\centering
\includegraphics[width=0.7\textwidth]{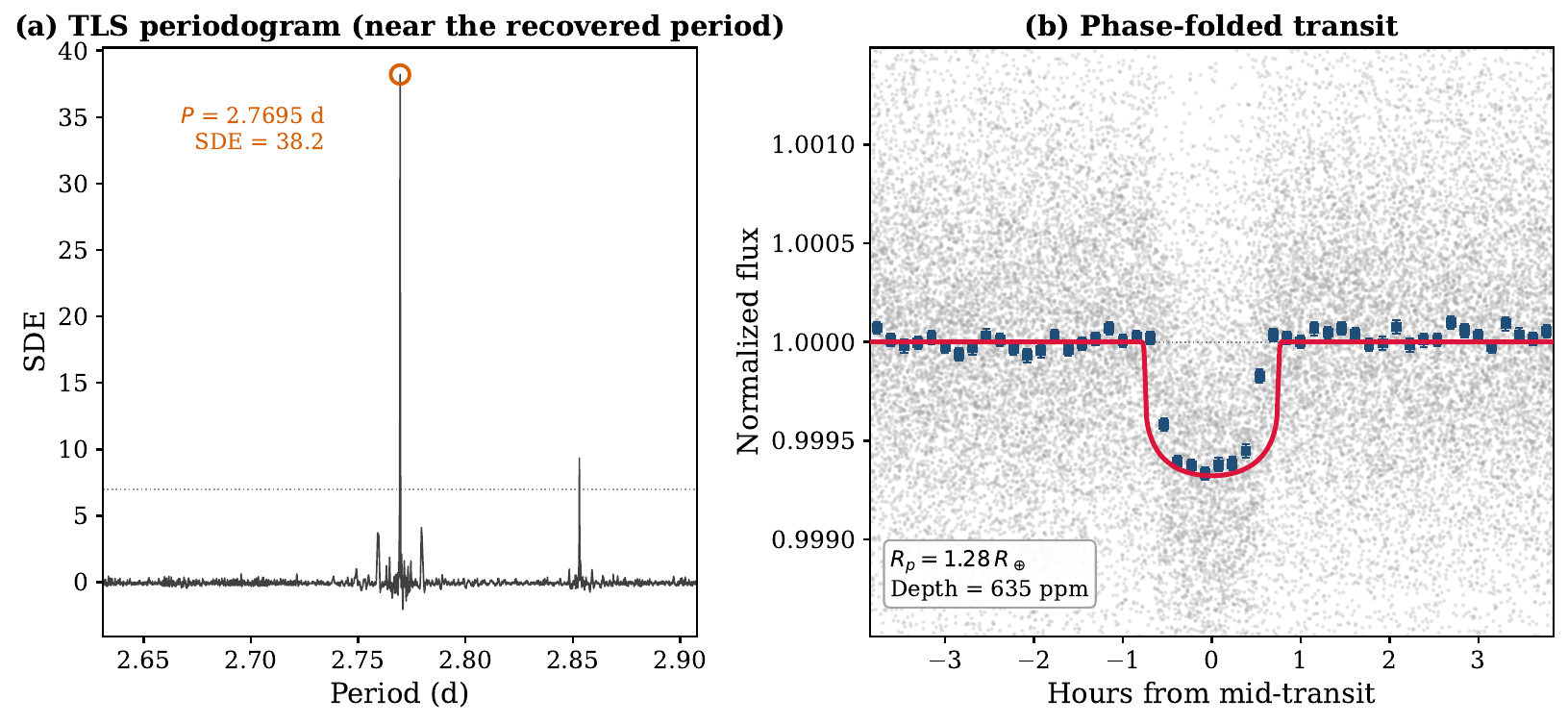}
\caption{Representative recovered confirmed planet, TOI-4599.01 (TIC~307809773; $P = 2.77$\,d, depth $635$\,ppm, $R_p \approx 1.3\,R_\oplus$). (a) The full-baseline TLS periodogram, peaking at $\mathrm{SDE} = 38.2$ at the recovered period. (b) The phase-folded transit with the limb-darkened model (format as in Fig.~\ref{fig:grid_recoveries}).}
\label{fig:showcase_recovery}
\end{figure}

Planet and candidate dispositions are taken from the ExoFOP-TESS snapshot of 2026 June 29; under this snapshot the fixed sample of 461 hosts carries 202 confirmed planets: 193 within the $0.5$--$100$\,d search range and nine below the $0.5$\,d grid floor (none lies beyond 100\,d). Both detection engines run on every target and their candidate lists are merged per host (Sect.~\ref{sec:segmented}). In the $0.5$--$100$\,d range the merged search recovers 165 of 193 in-range confirmed planets (85.5\%; 85/96 in M0--M2, 80/97 in M3--M6), the full-baseline and segmented engines contributing 28 and 23 unique recoveries beyond their 114 in common (Table~\ref{tab:cp_funnel}); all but one (TOI-5720.01, at $3P$) are at the archive period, matching the catalog to a median deviation of $0.0004\%$ (worst case $0.47\%$, TOI-2267.01). Eight of the nine confirmed planets below the $0.5$\,d floor are recovered through an in-range harmonic and restored to their catalog periods by the sub-floor arbitration of Sect.~\ref{sec:segmented} (sole loss TOI-6000.01), bringing the overall recovery to 173 of 202 confirmed planets (85.6\%). These recovery rates quantify the re-detection of the known TOI catalog and complement the blind injection--recovery completeness of \citet{Tschudi2026}, which maps the pipeline's detection sensitivity across period and radius. Figure~\ref{fig:showcase_recovery} illustrates a representative recovery, the confirmed planet TOI-4599.01.

One catalog planet, TOI-406.02, orbits near a $2{:}1$ period ratio with a sibling planet of deeper transit and is recovered by neither search; it is counted as not detected. In three further systems (TOI-1730.03, TOI-2095.01, TOI-4336\,A.01), a peak at the catalog period or its $2\times$ alias reaches $\mathrm{SDE_{raw}} \geq 7$ but falls below threshold after the red-noise correction (Table~\ref{tab:app_missed}).

\begin{deluxetable*}{lccc}
\tabletypesize{\small}
\tablecaption{Recovery and event-time screening of the ExoFOP confirmed planets (CP) with a period in the $0.5$--$100$\,d search range. Recovered $=$ at-archive $+$ harmonic; percentages are of the in-range CP.\label{tab:cp_funnel}}
\tablehead{\colhead{} & \colhead{M0--M2} & \colhead{M3--M6} & \colhead{Total}}
\startdata
ExoFOP CP in the search range\tablenotemark{a} & 96 & 97 & 193 \\
ExoFOP CP below the 0.5\,d floor          & 2 & 7 & 9 \\
Recovered, full-baseline search alone     & 70 & 72 & 142 (73.6\%) \\
Recovered, segmented search alone         & 68 & 69 & 137 (71.0\%) \\
\textbf{Recovered, merged (in range)}     & \textbf{85 (88.5\%)} & \textbf{80 (82.5\%)} & \textbf{165 (85.5\%)} \\
\quad by both searches                    & 53 & 61 & 114 \\
\quad full-baseline search only           & 17 & 11 & 28 \\
\quad segmented search only               & 15 & 8  & 23 \\
\quad \textit{of which, at an alias}      & 0  & 1  & 1 \\
Recovered below the floor (sub-floor arbitration) & 2 & 6 & 8 \\
Not detected                              & 11 & 17 & 28 \\
\cutinhead{Event-time screen on the recovered}
\quad \textsc{pass}                        & 81 & 76 & 157 \\
\quad \textsc{flag} / \textsc{flag\_lowsens} (retained) & 4 & 4 & 8 \\
\quad \textsc{fail}\,$\to$\,\texttt{ALIAS\_FALSE\_POSITIVE} & 0 & 0 & 0 \\
False positives removed among recovered CP (any stage) & 0 & 0 & 0 \\
\textbf{CP retained after screening}       & \textbf{85} & \textbf{80} & \textbf{165 (85.5\%)} \\
\enddata
\tablenotetext{a}{The nine sub-floor CP (Sect.~\ref{sec:recovery_tls}) are excluded here.}
\tablecomments{Per-planet causes in Table~\ref{tab:app_missed}.}
\end{deluxetable*}

Confirmed-planet recovery with the merged search is 88.5\% (85/96) in M0--M2 and 82.5\% (80/97) in M3--M6, the six-percentage-point decline toward the later types being expected from their fainter magnitudes and higher activity levels \citep{Tschudi2026}.

\subsection{Vetting Acceptance}
\label{sec:recovery_vetting}

Of the 165 recovered in-range confirmed planets (Sect.~\ref{sec:recovery_tls}), four are non-analyzable: three have no usable target pixel file on MAST (\texttt{INCONCLUSIVE}) and one single-sector host in a crowded field returns zero total probability from a degenerate flux-ratio configuration (\texttt{STRONG\_CANDIDATE\_VISUAL}). Among the 161 analyzable confirmed planets, 152 (94.4\%) receive a positive disposition (\texttt{LIKELY\_PLANET}, \texttt{PENDING}, or \texttt{VALIDATED}); the remaining nine are \texttt{AMBIGUOUS}: on-target signals whose conservative false-positive probability lies near the 50\% decision threshold (Sect.~\ref{sec:vetting}). No recovered confirmed planet receives a terminal false-positive, eclipsing-binary, or stellar-variability disposition, while on the same hosts the vetting removes 10 non-catalog signals as false positives and the event-time screen a further eight as window-function aliases. The pipeline emits the \texttt{VALIDATED} disposition on the campaign run for five of these confirmed planets; the rest remain at the candidate level, several meeting the $\mathrm{FPP} < 1.5\%$ and $\mathrm{NFPP} < 0.1\%$ thresholds but held below validation by the $R_p > 8\,R_\oplus$ cap (Sect.~\ref{sec:limitations}). Table~\ref{tab:cp_pipeline} traces the full pipeline from transit search through verification.

\begin{deluxetable}{lccc}
\tabletypesize{\footnotesize}
\tablecaption{Confirmed planet pipeline: transit search through verification.\label{tab:cp_pipeline}}
\tablehead{\colhead{} & \colhead{M0--M2} & \colhead{M3--M6} & \colhead{Total}}
\startdata
Recovered (merged)        & 85              & 80              & 165 \\
Non-analyzable            & 2               & 2               & 4 \\
Analyzable                & 83              & 78              & 161 \\
\textbf{Vetting accepted} & \textbf{78 (94.0\%)} & \textbf{74 (94.9\%)} & \textbf{152 (94.4\%)} \\
Ambiguous (on-target)     & 5               & 4               & 9 \\
Vetting rejected          & 0               & 0               & 0 \\
\cutinhead{Gaia~DR3 verification (147 candidates; 5 \textsc{validated}, held)}
PLANET\_CANDIDATE         & 57              & 57              & 114 \\
PC\_HEB\_CAUTION          & 5               & 4               & 9 \\
NEEDS\_HR\_IMAGING        & 14              & 10              & 24 \\
FALSE\_POSITIVE           & 0               & 0               & 0 \\
\enddata
\tablecomments{Non-analyzable $=$ \texttt{INCONCLUSIVE} or \texttt{STRONG\_CANDIDATE\_VISUAL}; vetting accepted $=$ \texttt{LIKELY\_PLANET}, \texttt{VALIDATED}, or \texttt{PENDING} (awaiting Gaia analysis).}
\end{deluxetable}

The 123 confirmed planets reaching \texttt{PLANET\_CANDIDATE} (114) or \texttt{PLANET\_CANDIDATE\_HEB\_CAUTION} (9) have background eclipsing binary scenarios excluded by Gaia~DR3. The 24 \texttt{NEEDS\_HR\_IMAGING} planets carry a Gaia-resolved neighbor (separations ${\ge}0.4\arcsec$; \citealt{Fabricius2021}) inside the photometric aperture, bright enough to host the signal, which the TESS aperture photometry does not separate; these include well-known planets confirmed through radial velocities (RV) and speckle imaging (e.g., TOI-1231\,b, TOI-732\,b); the flag marks a contaminated aperture to be cleared by high-resolution or on-target ground-based follow-up \citep{Ziegler2018}. No confirmed planet is localized off-target by the aperture-correlation test.

\subsection{False Negative Taxonomy}
\label{sec:false_negatives}

The 28 confirmed planets missed by the merged detection stage (11 M0--M2, 17 M3--M6; Table~\ref{tab:cp_funnel}) are not recovered at the catalog period; Table~\ref{tab:missed_causes} tallies the causes, with the per-planet catalog in Table~\ref{tab:app_missed}. Two regimes dominate and are the expected ones: no signal at any period, or a peak at the catalog period that remains below the detection threshold, for transits too shallow for the available baseline or masked by stellar activity. One peak is surfaced by the segmented search but fails its federation gates, the measured cost of the frozen detection thresholds (Sect.~\ref{sec:segmented}). Three structural causes account for the rest. Iterative masking extracts a stronger signal on the same host before the search reaches the planet (the two $2{:}1$ cases of Sect.~\ref{sec:recovery_tls} are its verified instances). Some hosts have photometry the pipeline cannot cleanly search: a diluted triple-star aperture, and an extremely young flare-active star (the regime characterized in the limitation tests of \citealt{Tschudi2026}). Finally, the $1.0$\,d Earth-rotation deadband rejects by construction. Some searched sectors carry only lower-cadence full-frame-image photometry, no 2-minute SPOC product existing for them, and the coarser sampling further reduces sensitivity to short, shallow transits.

\begin{deluxetable*}{lccc}
\tabletypesize{\small}
\tablecaption{Causes of non-detection for the 28 missed confirmed planets.\label{tab:missed_causes}}
\tablehead{\colhead{Cause} & \colhead{M0--M2} & \colhead{M3--M6} & \colhead{Total}}
\startdata
No signal at any period & 0 & 8 & 8 \\
Peak below the SDE threshold, raw or red-noise-corrected & 6 & 4 & 10 \\
Segmented peak failing the federation gates & 0 & 1 & 1 \\
Stronger signal extracted first (iterative masking)\tablenotemark{a} & 3 & 3 & 6 \\
Unsearchable host photometry\tablenotemark{b} & 2 & 0 & 2 \\
Instrumental deadband ($1.0$\,d Earth-rotation alias) & 0 & 1 & 1 \\
Total & 11 & 17 & 28 \\
\enddata
\tablenotetext{a}{For all six, a stronger signal on the same host is extracted before the search reaches the planet, indistinguishable from simple non-detection in the run summary; TOI-406.02 additionally sits at a $2{:}1$ ratio to its brighter sibling (Sect.~\ref{sec:recovery_tls}).}
\tablenotetext{b}{A triple-star aperture diluted by companions, and a $22$\,Myr star whose flares and spot modulation the detrending does not separate from transits.}
\tablecomments{Per-planet causes in Table~\ref{tab:app_missed}.}
\end{deluxetable*}

Full catalogs of recovered (Table~\ref{tab:app_recovered}) and missed planets (Table~\ref{tab:app_missed}) are provided in Appendices~\ref{app:recovered}--\ref{app:missed}.

\section{Planet candidate recovery}
\label{sec:pc_recovery}

Of the 311 PC signals in the sample (Table~\ref{tab:survey_overview}), 25 fall outside the search range ($P > 100$\,d, ultra-short, or no reported period) and are excluded, leaving 286 evaluable signals across 303 host targets. These are unconfirmed transit signals that have not been validated through radial velocity or statistical methods. Applying the validated pipeline provides an independent assessment of which candidates survive the full detection, vetting, and verification chain.

\subsection{Transit Search}
\label{sec:pc_tls}

Both detection engines run on every candidate host and their lists are merged per host (Sect.~\ref{sec:segmented}). In the $0.5$--$100$\,d range the merged search recovers 225 of 286 in-range planet candidates (78.7\%; 151/187 in M0--M2, 74/99 in M3--M6), the two engines contributing 63 and 59 unique recoveries beyond their 103 in common (Table~\ref{tab:pc_funnel}); of these, 221 are at the archive period and four at a half-period alias, three of them showing the deep morphology of an eclipsing binary (dispositioned two \texttt{LIKELY\_EB} and one \texttt{FALSE\_POSITIVE}, the fourth \texttt{LIKELY\_PLANET}). The 61 unrecovered signals are dominated by genuine non-detections (45, $74\%$: below the SDE threshold of 7, too shallow for the available baseline, activity-masked, or not producing a coherent periodic pattern); two (TOI-4572.01, TOI-7694.01) fall in the $13.7$\,d instrumental deadband (Sect.~\ref{sec:detection}), and the remaining 14 ($23\%$) are detected on the host but not at the cataloged period. Non-recovery alone is not evidence of a false positive, since genuine confirmed planets are missed for the same detection-limited reasons (below the SDE threshold or masked by activity; Sect.~\ref{sec:false_negatives}).

\begin{deluxetable*}{lccc}
\tabletypesize{\small}
\tablecaption{Recovery and event-time screening of the ExoFOP planet candidates (PC) with a period in the $0.5$--$100$\,d search range. Recovered $=$ at-archive $+$ alias, merged over the full-baseline and segmented searches; percentages are of the 286 in-range PC. Unlike the confirmed planets, the event-time screen is expected to remove some PC: an unconfirmed candidate may be a window-function alias.\label{tab:pc_funnel}}
\tablehead{\colhead{} & \colhead{M0--M2} & \colhead{M3--M6} & \colhead{Total}}
\startdata
ExoFOP PC in the search range & 187 & 99 & 286 \\
Recovered, full-baseline search alone     & 118 & 48 & 166 (58.0\%) \\
Recovered, segmented search alone         & 112 & 50 & 162 (56.6\%) \\
\textbf{Recovered, merged}                & \textbf{151 (80.7\%)} & \textbf{74 (74.7\%)} & \textbf{225 (78.7\%)} \\
\quad by both searches                    & 79 & 24 & 103 \\
\quad full-baseline search only           & 39 & 24 & 63 \\
\quad segmented search only               & 33 & 26 & 59 \\
\quad at the archive period               & 150 & 71 & 221 \\
\quad at an integer multiple/sub-multiple alias & 1 & 3 & 4 \\
Missed                                    & 36 & 25 & 61 \\
\quad no signal on host                   & 29 & 18 & 47 \\
\quad signal(s) on host, none at PC $P$   & 7 & 7 & 14 \\
\cutinhead{Event-time screen on the recovered}
\quad \textsc{pass}                        & 132 & 64 & 196 \\
\quad \textsc{flag} (retained)             & 9 & 5 & 14 \\
\quad \textsc{flag\_lowsens} (retained)    & 10 & 3 & 13 \\
\quad \textsc{fail}\,$\to$\,\texttt{ALIAS\_FALSE\_POSITIVE} & 0 & 1 & 1 \\
\quad no event-time verdict\tablenotemark{a} & 0 & 1 & 1 \\
False positives removed among recovered PC (any stage) & 6 & 5 & 11 \\
\textbf{PC retained after screening}       & \textbf{145 (77.5\%)} & \textbf{69 (69.7\%)} & \textbf{214 (74.8\%)} \\
\enddata
\tablenotetext{a}{The $0.43$\,d half-period alias, below the $0.5$\,d search floor, dispositioned \texttt{LIKELY\_EB} at the vetting stage.}
\end{deluxetable*}

The lower recovery rate compared to confirmed planets (78.7\% versus 85.5\%) reflects three factors: (1)~planet candidates include weaker signals that have not been independently confirmed, (2)~a fraction of ExoFOP planet candidates may be genuine false positives (eclipsing binaries, systematic artifacts) that the pipeline correctly does not recover, and (3)~the M3--M6 planet candidate sample has fewer available sectors (median 4 versus 5 for M0--M2), so that fewer transits are stacked, the per-signal detection statistic is lower, and a larger fraction of genuine signals falls below the SDE threshold.

\subsection{Vetting and Verification}
\label{sec:pc_vetting}

Of the 225 recovered candidates, one had already failed the event-time screen as a window-function alias (\texttt{ALIAS\_FALSE\_POSITIVE}; Sect.~\ref{sec:pc_tls}), and a further 36 are non-analyzable, the pixel-level analysis having no usable input (\texttt{INCONCLUSIVE}: missing target pixel file or WCS failure) or returning zero total probability from a degenerate multisector flux-ratio (\texttt{STRONG\_CANDIDATE\_VISUAL}), leaving 188 entering the vetting stage (Table~\ref{tab:pc_pipeline}); large-radius signals ($R_p > 8\,R_\oplus$) are vetted in full, their statistical validation alone capped at the candidate level. Of the 188 analyzable signals, 146 (77.7\%) receive a positive disposition, about 17 percentage points below the 94.4\% measured for confirmed planets, as expected if a fraction of the unconfirmed candidates are genuine false positives; the remaining 42 are 32 \texttt{AMBIGUOUS} signals (on-target detections whose conservative false-positive probability sits near the 50\% threshold, retained for follow-up) and ten terminal rejections (six \texttt{FALSE\_POSITIVE}, four \texttt{LIKELY\_EB}).

After Gaia~DR3 verification (Table~\ref{tab:pc_pipeline}), the final dispositions of the 143 vetting-accepted candidates entering verification (the three \texttt{VALIDATED} are held at the vetting stage) are: 87 \texttt{PLANET\_CANDIDATE}, 9 \texttt{PLANET\_CANDIDATE\_HEB\_CAUTION}, and 47 \texttt{NEEDS\_HR\_IMAGING}. No candidate is identified as a background eclipsing binary false positive at verification: where a Gaia-resolved neighbor (separations ${\ge}0.4\arcsec$; \citealt{Fabricius2021}) contributes flux to the photometric aperture that the TESS photometry does not separate, the signal is conservatively routed to \texttt{NEEDS\_HR\_IMAGING} for high-resolution follow-up \citep{Ziegler2018} rather than to a terminal rejection (Sect.~\ref{sec:vetting}).

\begin{deluxetable}{lccc}
\tabletypesize{\footnotesize}
\tablecaption{Planet candidate pipeline: transit search through Gaia~DR3 verification.\label{tab:pc_pipeline}}
\tablehead{\colhead{} & \colhead{M0--M2} & \colhead{M3--M6} & \colhead{Total}}
\startdata
\cutinhead{Transit search}
Evaluable PC               & 187             & 99              & 286 \\
Merged recovery            & 151             & 74              & 225 \\
Removed at the event-time screen        & 0               & 1               & 1 \\
Non-analyzable             & 24              & 12              & 36 \\
Analyzable                 & 127             & 61              & 188 \\
\cutinhead{Vetting}
Accepted                   & 107             & 39              & 146 \\
Ambiguous (on-target)      & 14              & 18              & 32 \\
Rejected (terminal)        & 6               & 4               & 10 \\
\cutinhead{Gaia~DR3 verification (143 candidates; 3 reach the validation threshold, held)}
\textbf{PLANET\_CANDIDATE} & \textbf{61}     & \textbf{26}     & \textbf{87} \\
PC\_HEB\_CAUTION           & 7               & 2               & 9 \\
NEEDS\_HR\_IMAGING         & 38              & 9               & 47 \\
FALSE\_POSITIVE            & 0               & 0               & 0 \\
\enddata
\end{deluxetable}

\subsection{Follow-up Categories for the Recovered Candidates}
\label{sec:followup}

The 87 signals reaching \texttt{PLANET\_CANDIDATE} status (Table~\ref{tab:pc_pipeline}) are split into two follow-up categories (Table~\ref{tab:followup_categories}):

\begin{itemize}
    \item Standard candidates (54 signals, $R_p < 8\,R_\oplus$, depth $< 10\,000$\,ppm): background eclipsing binary scenarios excluded by Gaia~DR3. Priority targets for radial velocity confirmation.
    \item Large-radius candidates (33 signals, $R_p \geq 8\,R_\oplus$ or depth $> 10\,000$\,ppm): statistical validation is not possible above $8\,R_\oplus$ \citep{Giacalone2021}, so these are dispositioned \texttt{LIKELY\_PLANET}. The observed depths ($10\,000$--$94\,000$\,ppm) are consistent with giant planets, brown dwarfs, or eclipsing binaries at a stellar or substellar mass transiting $R_\star \approx 0.3$--$0.5\,R_\odot$ hosts \citep[for context, the hot-Jupiter occurrence rate is $0.27 \pm 0.09\%$ for early M dwarfs;][]{Gan2023}. On the smallest hosts the depth cut also retains deep sub-Neptune transits (a $10\,000$\,ppm transit on a $0.3\,R_\odot$ star is only ${\sim}3.3\,R_\oplus$), so the bin is not exclusively giant. A literature and ellipsoidal-variation cross-check (Sect.~\ref{sec:candidates_revised}) identifies a fraction as eclipsing binaries or brown dwarfs; radial velocity mass measurements are required to separate the genuine giant planets from these. The deepest, TOI-5628.01 ($94\,000$\,ppm, $\mathrm{FPP} = 50.5\%$), formally meets the eclipsing-binary depth criterion (Appendix~\ref{app:flowchart}); it is retained here because its host lacks a measured radius, the default $0.4\,R_\odot$ being assumed, and awaits stellar characterization.
\end{itemize}

The nine \texttt{PLANET\_CANDIDATE\_HEB\_CAUTION} signals (host Gaia RUWE~$> 1.4$, indicating a possible unresolved companion) and the 47 \texttt{NEEDS\_HR\_IMAGING} signals (a Gaia-resolved neighbor inside the photometric aperture, bright enough to host the signal; Sect.~\ref{sec:recovery_vetting}) require high-resolution imaging or on-target ground-based photometry to exclude contaminants before radial velocity resources are committed \citep{Ziegler2018}. The ephemerides and full vetting-check set for the 87 recovered planet candidates are tabulated as a ground-based follow-up reference in Appendix~\ref{app:pc_phasefolds} (Table~\ref{tab:checks_pc}).

\begin{deluxetable*}{llcl}
\tabletypesize{\footnotesize}
\tablecaption{Planet candidate follow-up categories.\label{tab:followup_categories}}
\tablehead{\colhead{Category} & \colhead{Criteria} & \colhead{$N$} & \colhead{Follow-up}}
\startdata
Standard     & $R_p < 8\,R_\oplus$, depth $< 10\,000$\,ppm & 54 & Priority RV \\
Large-radius & $R_p \geq 8\,R_\oplus$ or depth $> 10\,000$\,ppm & 33 & RV (HJ versus EB) \\
HEB caution  & Host RUWE $> 1.4$              & 9  & Imaging $+$ RV \\
NHR          & Resolved Gaia neighbor         & 47 & HR imaging \\
\enddata
\tablecomments{HJ = hot Jupiter; EB = eclipsing binary; HEB = hierarchical eclipsing binary; NHR = \texttt{NEEDS\_HR\_IMAGING}.}
\end{deluxetable*}

\section{New transit candidates after alias-risk screening}
\label{sec:candidates_revised}

Beyond recovering the cataloged signals, the iterative search identifies detections not listed in ExoFOP. Each was passed through the event-time screen (Sect.~\ref{sec:screening}), then vetted (Sect.~\ref{sec:vetting}) and verified against the Gaia field (Sect.~\ref{sec:verification}); where post-detection photometry exists, a held-out-sector predicted-transit check is also reported.

After this screen, signals not attributable to a cataloged TOI are retained as new transit candidates, and several signals initially treated as independent detections are reclassified as recoveries of cataloged TOIs, four of them with their orbital periods determined here (Sect.~\ref{sec:recoveries_period}). One further retained signal, the confirmed planet TOI-237\,c (validated by \citealt{Timmermans2026}), is recovered blind and treated here as a recovery and screen calibration point rather than a new candidate. The screen retains 13 new transit candidates.

No retained signal reaches the joint statistical-validation thresholds of \citet{Giacalone2021} (FPP $< 1.5\%$, NFPP $< 0.1\%$). This reflects the conservative all-sector screen, not the candidates: the recovered confirmed planets mostly miss the same thresholds (only $11\%$ pass; median FPP $9.1\%$; Sect.~\ref{sec:limitations}) because the screen counts every resolved Gaia neighbor as a possible host. The FPP therefore ranks candidates and flags contaminated apertures rather than validating planets. The strongest candidates are instead cleared of their main false-positive channel (a blend or a stellar companion) with archival speckle imaging or reconnaissance spectroscopy, and reported as strong candidates; a firm validation still requires a radial-velocity mass and is left to dedicated follow-up.

\begin{figure}[!t]
\centering
\includegraphics[width=0.58\textwidth]{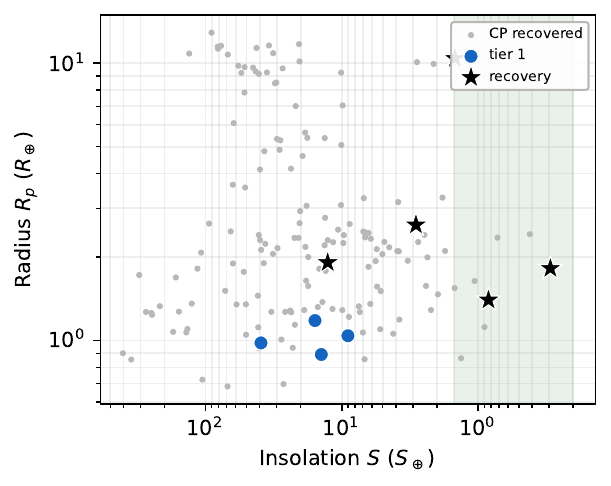}
\caption{Insolation--radius diagram of the recovered confirmed planets (gray), the four new tier~1 candidates (blue), and the five recovered cataloged TOIs whose orbital periods are determined or recovered here (black stars). The shaded band marks the optimistic habitable zone \citep{Kopparapu2013}, drawn for a representative host $T_{\rm eff}$ (per-star edges differ slightly); TOI-2433.01 and TOI-4353.01 each lie within their own optimistic zone. Insolation increases to the left. Highlighted radii are the values of Table~\ref{tab:ephemerides}; insolations are evaluated at the determined periods.}
\label{fig:population}
\end{figure}

The candidates are organized along a three-tier follow-up ladder, keyed to the detection and field state rather than the stellar-noise robustness of \citet{Tschudi2026}. Tier~1: a clean detection (folded dip detached and centered) with a clean field, the nearby false-positive probability (NFPP) of the conservative vetting (Sect.~\ref{sec:verification}) below $\sim$$10\%$ or the resolved-neighbor field cleared by archival high-resolution imaging; the residual on-target eclipsing binary is addressed by radial velocities or transit-shape analysis. Tier~2: a clean detection with an in-aperture Gaia neighbor bright enough to host the signal, awaiting imaging to localize the transit (\texttt{NEEDS\_HR\_IMAGING}), or one held short of tier~1 by a marginal statistic or grazing geometry. Tier~3: marginal detections, with SDE near the $7.0$ floor or a phase coherence in the stellar-variability regime. The recoveries of cataloged TOIs are treated first (Sect.~\ref{sec:recoveries_period}), including the blind co-recovery of the confirmed planet TOI-237\,c, which serves as a calibration point for the screen rather than a new candidate; Table~\ref{tab:screening_main} summarizes the detection and event-time screening of these recoveries and of the four new tier~1 candidates, the survey's strongest new signals, with the full ephemerides and the tier~2, tier~3, and non-retained signals collected in Appendix~\ref{app:candidate_tables} (Tables~\ref{tab:ephemerides} and~\ref{tab:candidates_screening}). Figure~\ref{fig:population} places the recoveries and the tier~1 candidates in the insolation--radius plane.

\subsection{Recoveries of Known TOIs: Period Determinations}
\label{sec:recoveries_period}

\begin{figure}[!tp]
\centering
\includegraphics[width=0.5\textwidth]{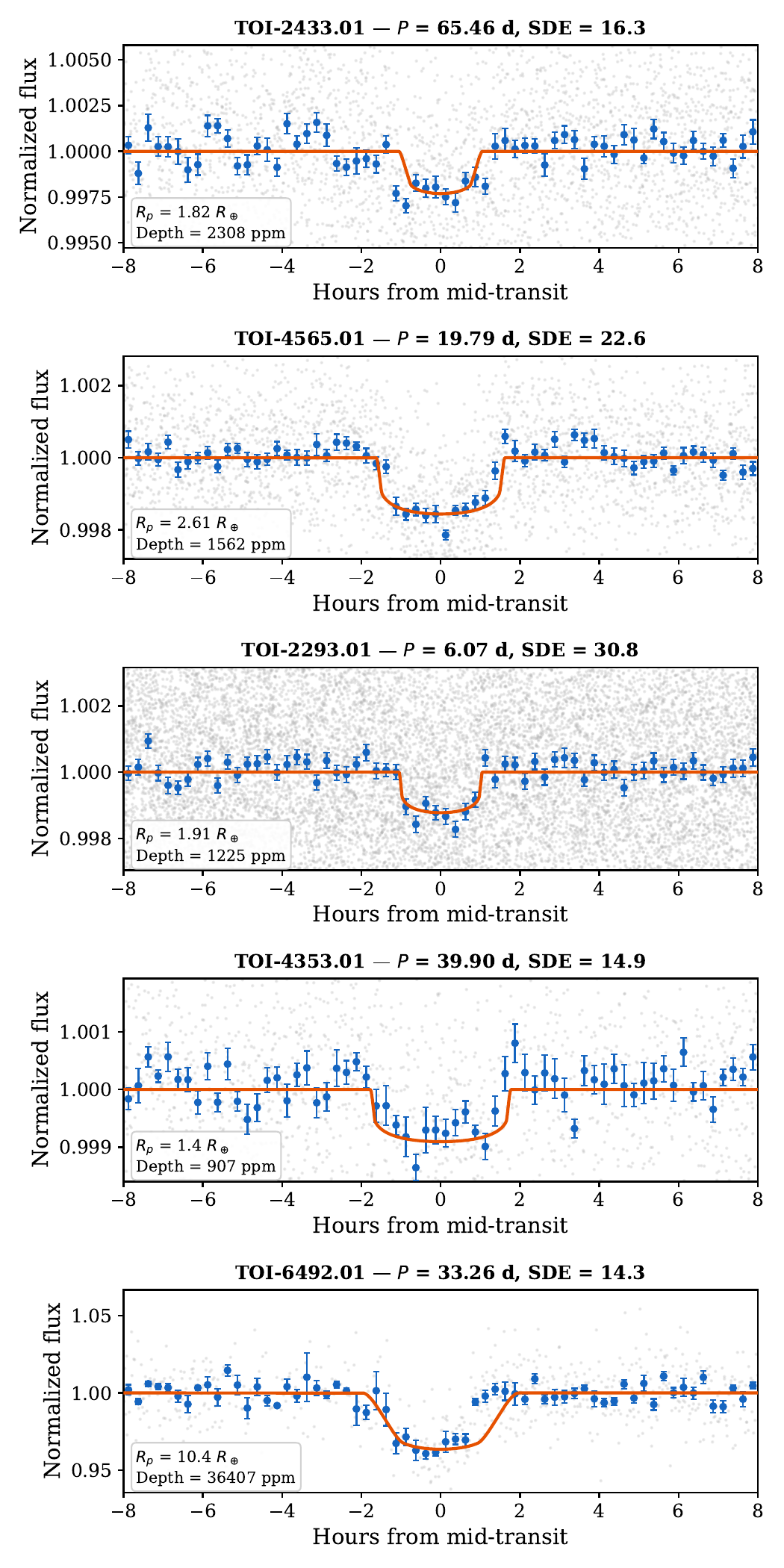}
\caption{Phase-folded TESS light curves of the five recovered cataloged TOIs at the periods determined or recovered here (Sect.~\ref{sec:recoveries_period}), each folded at that period, binned (blue circles with $1\sigma$ error bars) over the unbinned photometry (gray), with a limb-darkened transit model (orange) and the nominal planet radius and transit depth annotated.}
\label{fig:grid_recoveries}
\end{figure}

\candpar{TOI-2433.01.}
The longest-period signal of the survey ($P = 65.459$\,d, $\mathrm{SDE} = 16.3$, strict clock, seven events; Fig.~\ref{fig:grid_recoveries}), initially treated as an independent candidate, is the cataloged single-transit TOI-2433.01, which ExoFOP lists with an undefined period; the detection determines the orbital period. Four archival high-resolution imaging epochs (Robo-AO, Palomar, ShARCS, and WIYN speckle, none detecting a companion) confirm the empty field, while no radial velocities exist to exclude the on-target eclipsing binary that limits the conservative FPP (FPP $= 31\%$, NFPP $= 0$): the signal is reported as a strong candidate rather than a validated planet. With host parameters $T_{\rm eff} = 3323$\,K, $R_\star = 0.31\,R_\odot$, and $M_\star = 0.29\,M_\odot$, the candidate ($R_p = 1.82 \pm 0.21\,R_\oplus$, $S = 0.24 \pm 0.05\,S_\oplus$, $T_{\rm eq} = 178 \pm 9$\,K at $A = 0.3$) lies just outside the conservative outer edge and within the optimistic habitable zone \citep{Kopparapu2013}, the coolest signal of the survey. TESS sectors~119--121 (July--September 2027) will re-observe the target.

\candpar{TOI-4565.01.}
The $P = 19.786$\,d signal ($\mathrm{SDE} = 22.6$, strict clock, six events; Fig.~\ref{fig:grid_recoveries}), initially counted as a new candidate, coincides with the TFOPWG-revised ephemeris of TOI-4565.01 to within $2.7$\,s per cycle; the cataloged $692.515$\,d period, from two transits two years apart, is an exact $35{:}1$ alias. A rerun with the Gemini-8m speckle contrast curve leaves the nearby false-positive probability just above the validation threshold (FPP $= 0.13\%$, NFPP $= 0.12\%$ versus the $0.1\%$ threshold; Table~\ref{tab:screening_main}); a barycentric-frame cross-correlation of the eight archival CHIRON spectra performed here shows the systemic velocity stable to ${\sim}0.1$\,km\,s$^{-1}$ over 26 days, excluding a stellar-mass companion. With this channel cleared, the signal is a sub-Neptune-size strong candidate ($R_p = 2.61^{+0.16}_{-0.17}\,R_\oplus$). TESS sector~104 (observed May 2026), after the search data were frozen (March 2026), provides a direct out-of-sample test: at the single mid-transit time the archival ephemeris predicts within the sector (BTJD~$4189.41$, with BTJD $=$ BJD$-2\,457\,000$), a transit of depth ${\sim}1\,670$\,ppm with resolved ingress and egress is recovered, consistent with the expected $1\,562$\,ppm within the single-transit scatter, corroborating the ephemeris.

\candpar{TOI-2293.01.}
The cataloged TOI-2293.01 ($P = 6.068$\,d), recovered among the cataloged TOIs (Sect.~\ref{sec:pc_recovery}) at $\mathrm{SDE} = 30.8$ with a strict clock, is the strongest-supported recovery (Fig.~\ref{fig:grid_recoveries}). Its conservative FPP is on-target-eclipsing-binary-limited (FPP $= 6\%$, NFPP $= 0.03\%$); TRES reconnaissance spectroscopy over two epochs shows a single-lined M dwarf with no radial-velocity variation, excluding that channel, while Gemini-8m speckle imaging and ground-based nearby-eclipsing-binary photometry clear the resolved field. With both false-positive channels cleared by archival follow-up, the signal is reported as a strong candidate, formal validation awaiting a radial-velocity mass. A fixed-ephemeris refit with a stellar-density prior on $a/R_\star$ (the eccentric vetting MCMC did not converge and overshot the radius) gives $R_p = 1.91 \pm 0.12\,R_\oplus$, consistent with the cataloged $2.12 \pm 0.83\,R_\oplus$.

\candpar{TOI-4353.01.}
The $P = 39.899$\,d signal (Fig.~\ref{fig:grid_recoveries}) is a recovery of the cataloged duotransit TOI-4353.01, listed at $718.18$\,d, the baseline between its two cataloged transits: the search data now include two consecutive transits $39.899$\,d apart, one of them in the added sector~97, fixing the period directly as an exact $18{:}1$ alias of the cataloged $718.18$\,d duotransit baseline. Four transits over a $2\,577$\,d baseline hold a coherent linear ephemeris (residual $20.6$\,min), determining $P = 39.8991$\,d (Table~\ref{tab:ephemerides}). The conservative FPP is on-target-eclipsing-binary-limited (FPP $= 64\%$, NFPP $= 0$); CHIRON reconnaissance spectroscopy, two epochs, single-lined with a stable systemic velocity, excludes that channel, and SOAR speckle imaging confirms the empty field: a strong candidate cleared of both false-positive channels. The absolute-$K_s$ radius relation of \citet{Mann2015} gives $R_\star = 0.426 \pm 0.012\,R_\odot$, consistent with the TIC value and the Gaia parallax; with the $907$\,ppm transit depth this yields a near-Earth $R_p = 1.4 \pm 0.1\,R_\oplus$ and an insolation $S = 0.95 \pm 0.19\,S_\oplus$ that places the candidate in the optimistic habitable zone \citep{Kopparapu2013}, interior to the conservative inner edge, the survey's second such signal (with TOI-2433.01).

\candpar{TOI-6492.01.}
The cataloged duotransit TOI-6492.01 is recovered with its period determined at $P = 33.26$\,d, an exact $22{:}1$ alias of the cataloged $731.8$\,d inter-transit baseline, a deep ($3.6\%$) eclipse of a ${\sim}1\,R_{\rm Jup}$ object (Fig.~\ref{fig:grid_recoveries}). SPECULOOS and TRAPPIST photometry confirm the transit on-target and achromatic between the $R$ and $I{+}z$ bands, excluding a color-dependent blend, and Gemini-8m speckle bounds close companions. At this radius the transiting body is degenerate between a giant planet, a brown dwarf, and a very-low-mass star; only a radial-velocity mass, absent here, distinguishes them, and on an M dwarf a deep eclipse of this size is a priori more often a low-mass eclipsing binary than a rare giant planet. It is reported as a recovery with a determined period, its nature awaiting radial velocities, and, its cataloged $731.8$\,d period lying outside the search range, is not counted among the planet candidates.

\candpar{TOI-237\,c (confirmed planet; blind co-recovery).}
The $P = 1.74$\,d signal (shown in Fig.~\ref{fig:grid_tier1}) is recovered blind by both search engines at $\mathrm{SDE} = 18.7$, alongside the recovered host planet TOI-237\,b, and coincides with TOI-237\,c, a confirmed planet validated by \citet{Timmermans2026} from ground-based multicolor photometry while this analysis was in progress. It is therefore not a new candidate but a recovery of a known planet. Under the conservative all-sector screen this confirmed planet returns $\mathrm{FPP} = 40\%$ (its confirmed sibling, $5.5\%$), a direct calibration of the screen on an object independently known to be a planet; the blind recovery is an external check on the detection engines. It is reported here as a co-recovery and calibration point, and is not counted among the new candidates.

\subsection{Tier 1: Clean Detections in Clean Fields}
\label{sec:tier1}

\begin{figure}[!tp]
\centering
\includegraphics[width=0.5\textwidth]{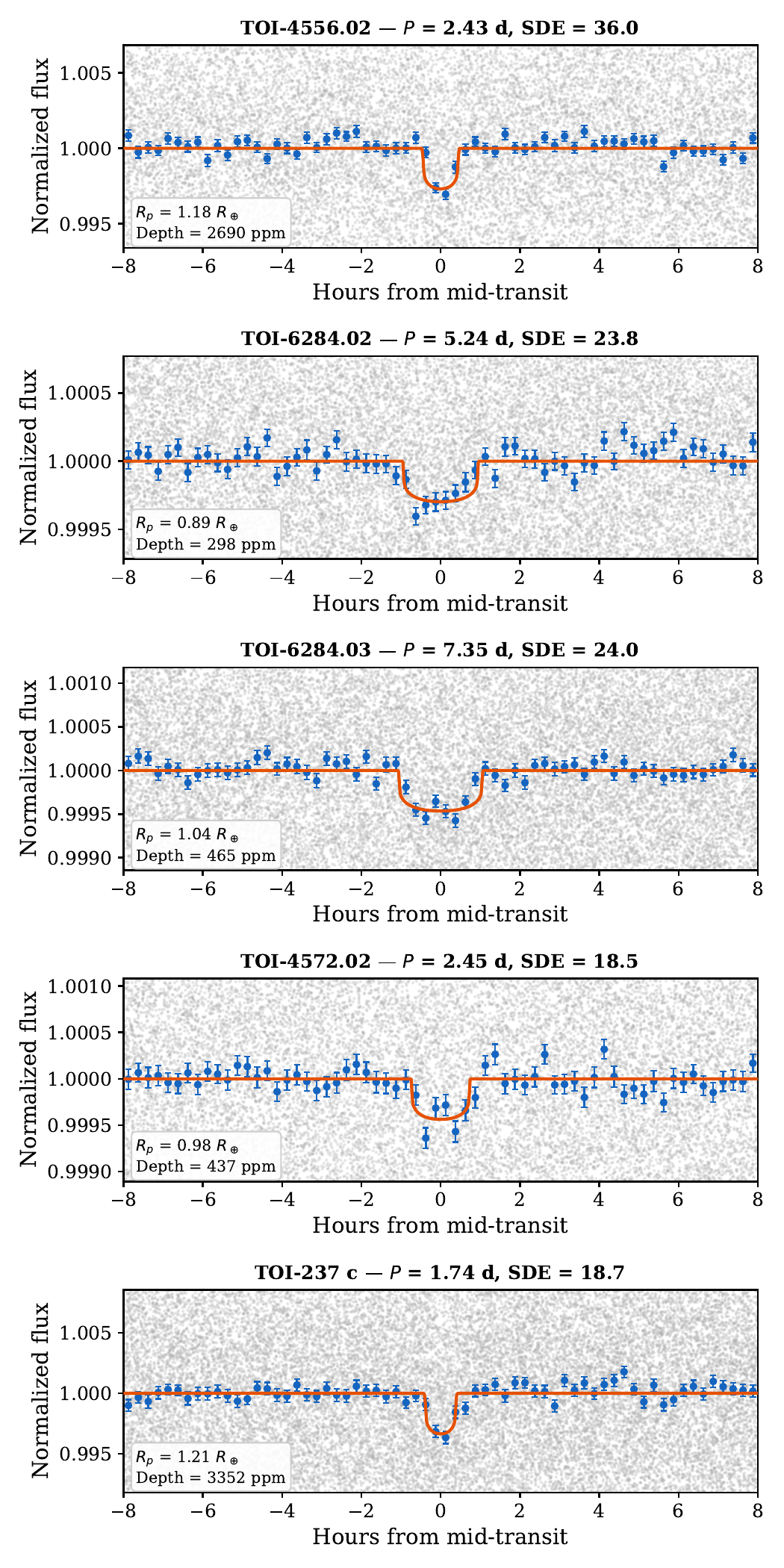}
\caption{Phase-folded TESS light curves of the four new tier~1 candidates and the confirmed planet TOI-237\,c (bottom), which is recovered blind and serves as a screen calibration point rather than a new candidate (Sect.~\ref{sec:tier1}, \ref{sec:recoveries_period}); format as in Fig.~\ref{fig:grid_recoveries}. Gray: unbinned photometry; blue circles: binned data with $1\sigma$ error bars; orange: limb-darkened transit model.}
\label{fig:grid_tier1}
\end{figure}

\candpar{TOI-4556.02.}
The $P = 2.43$\,d signal (Fig.~\ref{fig:grid_tier1}), detected blind at $\mathrm{SDE} = 36.0$ with a strict clock (44 events) alongside the cataloged TOI-4556.01, is the strongest new candidate of the survey. It is the only candidate with a true out-of-sample test, sectors~90 and 101 having been observed only after the search data were frozen: these held-out sectors contain 18 predicted epochs whose anchored stack reaches $7.2\sigma$, the significance expected of a genuine train of this depth. The conservative false-positive probability sits just above the validation threshold (FPP $= 2.46\%$, NFPP $= 2.40\%$; Table~\ref{tab:screening_main}), dominated by Gaia-resolved neighbors that on-target ground photometry would clear. The candidate is tier~1, and TESS sector~114 (February 2027) will provide a further out-of-sample test.

\candpar{TOI-6284.02 and TOI-6284.03.}
The bright host TOI-6284 ($T_{\rm mag} = 9.7$) carries, beside its recovered cataloged candidate TOI-6284.01 ($P = 3.45$\,d), two new strict-clock signals detected blind (Fig.~\ref{fig:grid_tier1}): a $P = 5.24$\,d signal, the shallowest new candidate of the survey ($\mathrm{SDE} = 23.8$, 33 events), and a $P = 7.35$\,d signal ($\mathrm{SDE} = 24.0$, 26 events). Gemini-8m speckle imaging excludes close companions and a ground-based nearby-eclipsing-binary check clears all 102 field neighbors to $2.5\arcmin$ at the transit depth; the residual channel for both is the on-target eclipsing binary, to be excluded by radial velocities, the tabulated NFPP of the $7.35$\,d signal ($19.2\%$) still counting the wide neighbors the ground check has cleared. Both candidates are placed in tier~1 on the cleared field. The three signals form a compact near-commensurable chain (period ratios $1.52$ and $1.40$, near $3{:}2$ and $7{:}5$), dynamically stable for masses from the nominal mass--radius relation \citep{ChenKipping2017}: the mutual-Hill separations of $20.5$ and $11.4$ clear the two-body limit \citep{Gladman1993} and their sum the three-planet boundary of \citet{Fabrycky2014}, the multiplicity and stability consistent with a planetary interpretation \citep{Lissauer2011}.

\candpar{TOI-4572.02.}
The $P = 2.45$\,d signal (Fig.~\ref{fig:grid_tier1}) is detected blind by both search engines at $\mathrm{SDE} = 18.5$, a centered, detached fold on a flat baseline; the event-time screen returns \textsc{flag}, at the per-event sensitivity limit for so shallow a signal. The cataloged TOI-4572.01 falls in the $13.7$\,d TESS window deadband and is not recovered, leaving this as the only detection on the host. TRES reconnaissance spectroscopy, six epochs over 352 days showing a single-lined spectrum with no radial-velocity variation, excludes the on-target eclipsing binary, and WIYN and SAI speckle imaging detect no companion. With the on-target eclipsing binary excluded and no companion in the field, the signal is placed in tier~1.

\begin{deluxetable*}{lcccccc}
\tabletypesize{\scriptsize}
\tablecaption{Detection and event-time screening of the recovered cataloged TOIs (including the confirmed planet TOI-237\,c, recovered blind) and the four new tier~1 candidates, the survey's strongest new signals. The full ephemerides and parameters (Table~\ref{tab:ephemerides}), and the tier~2, tier~3, and non-retained signals (Table~\ref{tab:candidates_screening}), are collected in Appendix~\ref{app:candidate_tables}; per-signal detail is given in the text.\label{tab:screening_main}}
\tablehead{\colhead{Designation} & \colhead{$P$ (d)} & \colhead{SDE} & \colhead{$p_{\rm noclock}$} & \colhead{FPP} & \colhead{NFPP} & \colhead{Outcome}}
\startdata
\cutinhead{Recovered cataloged TOIs (incl.\ the confirmed planet TOI-237\,c)}
TOI-2433.01 & 65.459  & 16.3 & $<0.005$ (PASS) & $31\%$   & $0$      & Rec. \\
TOI-4565.01 & 19.786  & 22.6 & $<0.005$ (PASS) & $0.13\%$ & $0.12\%$ & Rec. \\
TOI-2293.01 & 6.068   & 30.8 & $<0.005$ (PASS) & $6\%$    & $0.03\%$ & Rec. \\
TOI-4353.01 & 39.899  & 14.9 & FLAG\_LOWSENS  & $64\%$   & $0$      & Rec. \\
TOI-6492.01 & 33.264  & 14.3 & FLAG\_LOWSENS  & $72\%$   & $28\%$   & Rec. \\
TOI-237\,c  & 1.7449  & 18.7 & $<0.005$ (PASS) & $40.0\%$ & $0\%$    & Rec. (CP)\tablenotemark{a} \\
\cutinhead{New tier-1 candidates}
TOI-4556.02 & 2.4348  & 36.0 & $<0.005$ (PASS) & $2.46\%$ & $2.40\%$ & tier~1 \\
TOI-6284.02 & 5.2445  & 23.8 & $<0.005$ (PASS) & $6.7\%$  & $5.8\%$  & tier~1 \\
TOI-6284.03 & 7.3494  & 24.0 & $<0.005$ (PASS) & $24.9\%$ & $19.2\%$ & tier~1 \\
TOI-4572.02 & 2.4496  & 18.5 & $0.27$ (FLAG)  & $4.2\%$  & $1.9\%$  & tier~1 \\
\enddata
\tablenotetext{a}{TOI-237\,c is a confirmed planet \citep{Timmermans2026}, recovered blind; its FPP is a screen calibration point, and it is not counted among the new candidates.}
\tablecomments{The $p_{\rm noclock}$ verdict scale and \texttt{FLAG\_LOWSENS} are defined in Sect.~\ref{sec:screening}; SDE is the blind or windowed TLS statistic; FPP and NFPP are the all-sector median values (Sect.~\ref{sec:vetting}).}
\end{deluxetable*}

\subsection{Tier 2: Clean Detections Awaiting High-resolution Imaging}
\label{sec:tier2}

\candpar{TOI-2285.02.}
The $P = 9.67$\,d signal (Fig.~\ref{fig:grid_tier23}), detected blind at $\mathrm{SDE} = 20.3$ and passing the screen (23 events), is the strongest tier~2 candidate. \citet{Fukui2025}, revising the host's planet~b to $13.64$\,d, independently report a $9.67$\,d companion candidate, so the blind detection here co-detects that signal from an independent analysis, an external corroboration rather than a shared-photometry selection effect, although both detections rest on the same photometric source. The NFPP is carried by 36 resolved in-aperture Gaia neighbors rather than an unresolved close companion, so the high conservative FPP, unstable across the choice of sectors in so dense a field, reflects the field rather than the transit. The candidate is tier~2, awaiting high-resolution imaging to localize the transit.

\candpar{Further tier-2 candidates.}
Six further signals are clean detections awaiting the localization of the transit among the Gaia field stars (\texttt{NEEDS\_HR\_IMAGING}). TOI-1467.02 ($P = 3.15$\,d, Fig.~\ref{fig:grid_tier23}) lies near a $2{:}1$ ratio with the confirmed planet TOI-1467\,b, dynamically plausible for a real companion, but is grazing, with a radius left poorly constrained by the depth--impact-parameter degeneracy. TOI-2293.02 ($P = 35.4$\,d), a second candidate on the TOI-2293 host (Sect.~\ref{sec:recoveries_period}), pairs a clean, detached fold and TRES spectroscopy excluding the on-target eclipsing binary with only four timeable events. TOI-4353.02 ($P = 2.93$\,d) is a grazing, low-significance detection in the cleanest field of the group, the host single-lined in CHIRON, a faint in-aperture neighbor nonetheless keeping the disposition \texttt{NEEDS\_HR\_IMAGING}. TOI-4508.02 ($P = 2.28$\,d) lies near a $3{:}2$ ratio with the recovered TOI-4508.01. TOI-6578 carries two further candidates beside its recovered $2.754$\,d cataloged candidate, a compact multiplanet candidate system whose sparse archival follow-up leaves the field unresolved.

\subsection{Tier 3 and the Threshold-level Population}
\label{sec:tier3}

Two marginal detections are retained in tier~3. The $P = 20.3$\,d signal toward TOI-4360 (Fig.~\ref{fig:grid_tier23}), detected blind at $\mathrm{SDE} = 7.3$ just above the floor, passes the screen (12 events), is centered, on-target, and not a rotation harmonic; a SOAR speckle re-vetting lowers the background-eclipsing-binary probability tenfold, the residual an on-target eclipsing binary no radial velocities exclude. The candidate is temperate ($T_{\rm eq} \approx 255$--$280$\,K, $S \approx 1.0$--$1.4\,S_\oplus$ at $A = 0.3$) but rests on a weak detection with a non-flat out-of-transit baseline and TIC stellar parameters, is designated TOI-4360.02, and is reported with that caveat; an extended baseline is the decisive test. The second is the $P = 0.83$\,d ultra-short-period signal TOI-4349.02 (TIC~313889049, $\mathrm{SDE} = 9.3$; Fig.~\ref{fig:grid_tier23}), surfaced by the segmented search on a field cleared by Gemini speckle imaging but shallow and eclipsing-binary-limited ($\mathrm{FPP} = 31.6\%$), its folded dip neither detached nor centered and its per-event timing only flagging; it is reported at tier~3, not as a clean candidate.

\subsection{Withdrawn, Demoted, and Indeterminate Signals}
\label{sec:withdrawn}

The candidate signals that fail the final adjudication, none a cataloged TOI (Sect.~\ref{sec:pc_recovery} counts those rejections), are dispositioned in Table~\ref{tab:candidates_screening}. Three pass the timing screen but are rejected on physical grounds: the grazing $3.247$\,d signal toward TOI-2495 ($b = 0.97$, fold not detached) and the two nearby eclipsing binaries toward TOI-4559 ($9.277$ and $10.308$\,d, prob$_{\rm BEB} = 1.000$, $\mathrm{FPP} = \mathrm{NFPP} = 100\%$). Two more are excluded before the screen: the published false positive TOI-6255.02, identified by \citet{Dai2024} from a $27\arcsec$ centroid offset below the TRICERATOPS pixel resolution, and a high-order ($12{:}1$) alias of a cataloged ultra-short-period planet.

Three signals resist a terminal disposition. On TOI-663, an $11.892$\,d signal never crosses the blind detection floor ($\mathrm{SDE} \le 5.6$, within $4\%$ of the stellar rotation), and a quasi-periodic Gaussian-process refit of the 69 published radial velocities \citep{Cointepas2024} bounds any companion at this period to $K < 3.07$\,m\,s$^{-1}$ ($M_p \sin i < 7\,M_\oplus$, $95\%$): it is classified as indeterminate. A $54.34$\,d second signal on TOI-4360 is likewise demoted. On TOI-4349, a $31.4934$\,d signal passes the timing screen, but the cataloged duotransit TOI-4349.01 ($P = 749.1605$\,d) aliases to $31.2150$\,d, $0.88\%$ away and phasing both SPOC transits, and the blind search recovers neither period; the signal is demoted and TOI-4349.01 remains an unresolved duotransit. TESS sector~107 (August 2026) re-observes the target and will re-test both solutions.

\section{Discussion}
\label{sec:discussion}

\subsection{Comparison with Other M-dwarf Transit Surveys}
\label{sec:comparison}

Several systematic searches have targeted TESS M dwarfs. SHERLOCK \citep{Pozuelos2020, DevoraPajares2024} runs an iterative TLS search with vetting and statistical validation, but one host at a time; the QLP team produced the bulk of M-dwarf TOIs but is detection-first, without integrated Gaia background analysis; and LEO-Vetter \citep{Kunimoto2025} vets downstream of a TLS or box least squares \citep[BLS;][]{Kovacs2002} search ($93\%$ acceptance on pre-detected TOIs) but does not search. The difference here is scale: detection, vetting, and Gaia-based verification are applied uniformly across all 461 hosts rather than one target at a time, with the segmented semi-coherent search (Sect.~\ref{sec:segmented}) added for the gapped multisector baselines, complementing the uniform mid-to-late M-dwarf occurrence census of \citet{Gillis2026}. The new candidates have radii of $0.7$--$2.0\,R_\oplus$, consistent with the small-planet dominance around M dwarfs \citep{Dressing2015, Ment2023}.

The segmented, semi-coherent search introduced here (Sect.~\ref{sec:segmented}) differs in construction from existing multisector detection strategies: from the per-event single-event-statistic to multiple-event-statistic (SES-to-MES) combination of the Kepler and TESS SPOC pipelines \citep{Jenkins2002, Tenenbaum2013}, whose absolute per-event statistic is structurally insensitive to the window-function forest but requires individually significant events; from duotransit period matching \citep{Cooke2021}, which federates individually detectable transits rather than sub-threshold periodograms; and from deliberately campaign-independent searches \citep{Zink2021}, which merge only at catalog level. Federating the block periodograms themselves recovers the coherent stacking gain of the full baseline while keeping each detection statistic's noise floor local to its block.

\subsection{Archival Guardrails}
\label{sec:screening_lessons}

Three archival guardrails, applicable before any radial-velocity or ground-based follow-up, generalize from this survey's screening experience. First, when a detection matches an independently known period, require its predicted transit times to align, not the period value alone \citep{Dawson2010}: a coincidental period match can stack incoherent events into a false train, which only per-event timing separates from a genuine clock. Second, require the signal to reproduce in new data: several candidates here (TOI-4565.01, TOI-4556.02; Sect.~\ref{sec:candidates_revised}) were recovered again in TESS sectors observed after the search dataset was frozen, an out-of-sample confirmation stronger than any test on the search data alone. Third, treat a ground-based false-positive verdict as specific to its ephemeris: an eclipsing binary excluded at one period does not clear a neighboring one, each of which must be checked on its own transit times.

\subsection{Known Limitations}
\label{sec:limitations}

Ten systematic limitations are identified:

\begin{enumerate}\setlength{\itemsep}{0pt}\setlength{\parsep}{0pt}
    \item Validation capped above $8\,R_\oplus$: TRICERATOPS cannot distinguish planets from brown dwarfs or low-mass stars for $R_p > 8\,R_\oplus$ \citep{Giacalone2021}, so these are capped at the candidate level (radial-velocity masses required).

    \item Validation withheld on high-scatter folds: on a many-sector fold the per-bin error shrinks but an in-transit mismatch (transit-timing variations, residual period error) does not, saturating the false-positive probability of an on-target planet (on TOI-700, planet e saturates in 17 of 20 Monte Carlo draws while planet d folds cleanly; \citealt{Gilbert2023}); this is contained by taking the median of 20 draws, bounding the pre-fold period refinement to a coherence window \citep{Ofir2014}, and flagging a saturated draw distribution as degenerate.

    \item BIC on active M dwarfs: the Bayesian information criterion ($\Delta\mathrm{BIC}$) transit test assumes independent data points, but correlated stellar noise reduces the effective sample size \citep{Pont2006} and can misclassify a real transit as stellar variability or a spurious eclipsing-binary secondary. The red-noise floor and binned-error FPP correction of Sect.~\ref{sec:vetting} recover these signals, so no confirmed planet is rejected (Table~\ref{tab:cp_pipeline}); the test nonetheless stays degenerate on the most active hosts.

    \item In-aperture neighbors and conservative NFPP: Gaia~DR3 cannot resolve companions closer than $0.4\arcsec$ \citep{Fabricius2021}; a resolved neighbor contributing aperture flux is kept in the TRICERATOPS scenario set rather than cleared by the SPOC centroid alone (Sect.~\ref{sec:vetting}), since no survey clears a non-empty NFPP from a TESS difference image without on-target follow-up (\citealt{Giacalone2022, Barkaoui2024b} used ground-based or multicolor photometry). The NFPP is thus a conservative upper bound: even the recovered confirmed planets, with well-behaved FPP (median $9.1\%$), mostly miss the joint criterion of \citet{Giacalone2021} on TESS alone (only $11\%$ pass, the NFPP being the binding constraint), and an elevated RUWE flags a possible unresolved companion on nine recoveries and nine candidates (\texttt{PC\_HEB\_CAUTION}). Failing the criterion flags an in-aperture neighbor for high-resolution imaging (\texttt{NEEDS\_HR\_IMAGING}), not a false positive.

    \item MCMC impact-parameter degeneracy: the $b$--$R_p/R_\star$ degeneracy \citep{Espinoza2018} lets the MCMC settle at $b \approx 1$ for shallow transits, biasing the radius but not causing false rejections (Sect.~\ref{sec:vetting}).

    \item Iterative masking in multiplanet systems: masking a dominant signal before searching for the next can remove a shallower planet whose transits overlap it, as in the $2{:}1$ configuration of TOI-406.02; such losses leave no peak and are counted as not-detected (Table~\ref{tab:app_missed}). Simultaneous multisignal fitting would avoid them but is not implemented.

    \item Frozen federation thresholds in the segmented search: the segmented search (Sect.~\ref{sec:segmented}) recovers signals that a full-baseline scan dilutes on heavily gapped multiyear coverage, but it federates block detections against fixed thresholds, so a genuine sub-threshold peak at the catalog period can fall short of federation and be missed (Sect.~\ref{sec:false_negatives}).

    \item Stellar parameters from TIC only: all stellar radii, temperatures, and gravities are from the TESS Input Catalog \citep{Stassun2019} without spectroscopic verification. For the coolest M dwarfs the TIC $R_\star$ systematics can exceed the ${\sim}3\%$ of dedicated calibrations \citep{Mann2015} and propagate into the planet properties: a $10\%$ error on $R_\star$ gives ${\sim}10\%$ on $R_p$ and ${\sim}20\%$ on insolation ($S \propto R_\star^2$). Spectroscopy of the habitable-zone hosts is needed to refine these.

    \item Selection bias at the detection threshold: a signal selected by threshold crossing overstates its own detection statistic (the winner's curse), so a blind detection peak is not a measured period until it passes the event-level timing test calibrated on injected twins; statistical validation does not substitute for that screen (Sect.~\ref{sec:screening_lessons}).

    \item Pixel-level vetting needs a SPOC target pixel file: TRICERATOPS and the transit fit operate on the TESS pixel data, so a signal whose host has only Quick-Look or full-frame-image light curves, with no SPOC or TESS-SPOC target pixel file on MAST, cannot be vetted at the pixel level and is reported as inconclusive: 24 signals on 22 hosts ($6.2\%$ of the recovered signals). Combining the archival light curve with a full-frame-image cutout would recover them, and is left to future work.

    \end{enumerate}

\section{Conclusions}
\label{sec:conclusions}

The principal results of this work are:

\begin{enumerate}\setlength{\itemsep}{0pt}\setlength{\parsep}{0pt}
    \item A uniform end-to-end survey: this work reports the first homogeneous detection, vetting, and verification run, to the author's knowledge, on the 461 active ExoFOP M-dwarf TOI hosts. The merged search recovers 165 of the 193 in-range confirmed planets (85.5\%), of which 147 (76.2\%) reach a Gaia-verified candidate-level disposition at a 94.4\% vetting acceptance, with no confirmed planet flagged as a false positive; the 28 misses trace to stellar activity, survey-design limits, and iterative masking.

    \item An event-level timing screen: a per-event no-clock statistic, $p_{\rm noclock}$, is added to the statistical vetting to separate coherent transit trains from window-stacked noise (Sect.~\ref{sec:screening}). Calibrated on the window-alias false positive TOI-521\,c (\textsc{fail}) against 16 confirmed signals (15 \textsc{pass}, 1 \textsc{flag}, none \textsc{fail}), it makes the period of a blind $\mathrm{SDE} = 7$--$14$ peak a measurement only once it is passed, at archival cost and before any follow-up is committed.

    \item Planet-candidate recovery: of the 286 evaluable ExoFOP planet candidates, 225 (78.7\%) are recovered by the search. The full chain then yields 87 planet candidates (54 standard, $R_p < 8\,R_\oplus$, for radial-velocity confirmation, and 33 large-radius, requiring masses), 47 signals flagged for high-resolution imaging, and 9 with an elevated host RUWE; their ephemerides and vetting checks are tabulated for community follow-up (Appendix~\ref{app:pc_phasefolds}).

    \item New candidates and cataloged recoveries: the screen retains 13 new transit candidates across three confidence tiers. The two habitable-zone objects both lie in the optimistic zone and are cataloged recoveries whose period is determined here: TOI-2433.01, the coolest signal of the survey ($T_{\rm eq} = 178$\,K), its period determined here from seven recovered transits of a formerly single-transit TOI, and TOI-4353.01 ($718$\,d duotransit resolved to $39.90$\,d); TOI-4565.01 (a $35{:}1$ alias of $19.79$\,d) and the deep-eclipse TOI-6492.01 ($33.26$\,d) are likewise recovered. The most notable new system is a candidate three-planet chain on the bright TOI-6284: two new small candidates join the recovered cataloged candidate in a near-commensurable, dynamically stable configuration. The survey also blind-recovers TOI-237\,c, validated independently by \citet{Timmermans2026}, an external check on the detection.

    \item Conservative validation and follow-up: none of the cataloged-TOI recoveries or new candidates of Sect.~\ref{sec:candidates_revised} meets the joint validation thresholds ($\mathrm{FPP} < 1.5\%$, $\mathrm{NFPP} < 0.1\%$) under the all-sector conservative screen, which flags a Gaia neighbor bright enough to host the signal for high-resolution imaging rather than rejecting it as a false positive. The survey therefore validates no new planet: the five \texttt{VALIDATED} dispositions of Sect.~\ref{sec:recovery_vetting} are already-confirmed planets. The strongest recoveries and candidates are instead cleared of their dominant false-positive channel by archival follow-up (TRES and CHIRON reconnaissance spectroscopy, speckle imaging) and reported as strong candidates, formal validation deferred to dedicated observations. The new candidates will be registered as community TOIs on ExoFOP with the ephemerides of Table~\ref{tab:ephemerides}; and as TESS keeps revisiting these hosts, the survey's reach will extend toward the longer periods where the habitable zone of an M dwarf lies.
\end{enumerate}

\section*{Data availability}

Full catalogs of the 165 recovered and 28 missed confirmed planets are given in Appendices~\ref{app:recovered} and~\ref{app:missed}; the candidate ephemerides and screening dispositions are in Appendix~\ref{app:candidate_tables} (Tables~\ref{tab:ephemerides} and~\ref{tab:candidates_screening}). Machine-readable versions of the recovered-planet, missed-planet, and follow-up-reference catalogs are provided in the electronic edition. Pipeline code and candidate data are at \url{https://gitlab.com/yohanntschudi/exoplanets-project}. TESS photometry was obtained from the Mikulski Archive for Space Telescopes (MAST) at STScI (\url{https://mast.stsci.edu}). The TESS-SPOC full-frame-image light curves \citep{Caldwell2020} and the QLP light curves \citep{Huang2020a, Huang2020b} used here are available at MAST via \dataset[doi:10.17909/t9-wpz1-8s54]{https://doi.org/10.17909/t9-wpz1-8s54} and \dataset[doi:10.17909/t9-r086-e880]{https://doi.org/10.17909/t9-r086-e880}, respectively; the SPOC 2-minute light curves were retrieved from the same archive.

\begin{acknowledgments}
Analysis was performed using Python~3.11 with \texttt{lightkurve}~2.4, \texttt{transitleastsquares}~1.0.31, \texttt{triceratops}, \texttt{celerite2}~0.3, \texttt{emcee}~3.1, and \texttt{batman}~2.4. \texttt{triceratops} was used at development commit \texttt{8cd3633} (2026 April 9) rather than a packaged release: it corrects two numerical issues, reported through the package's issue tracker, that materially affect multisector false-positive probabilities. Pipeline computations were performed on a 32-vCPU server and on Apple M-series hardware.

The author thanks Ren\'e Heller for discussions on transit detection and false positive vetting.

AI transparency: pipeline code was developed with assistance from Claude (Anthropic) for code generation and debugging. All AI-generated code was reviewed, tested, and validated by the author. The manuscript text was written by the author; AI tools were used for language editing only. The full development history is preserved in the version control repository.

This paper includes data collected with the TESS mission, obtained from the MAST data archive at the Space Telescope Science Institute (STScI). Funding for the TESS mission is provided by the NASA Explorer Program. STScI is operated by the Association of Universities for Research in Astronomy, Inc., under NASA contract NAS~5-26555. This work has made use of data from the European Space Agency (ESA) mission Gaia, processed by the Gaia Data Processing and Analysis Consortium (DPAC). This research has made use of the NASA Exoplanet Archive and the Exoplanet Follow-up Observation Program (ExoFOP; \dataset[DOI: 10.26134/ExoFOP5]{https://doi.org/10.26134/ExoFOP5}), which are operated by the California Institute of Technology, under contract with the National Aeronautics and Space Administration under the Exoplanet Exploration Program.
\end{acknowledgments}

\facilities{TESS, Gaia, Exoplanet Archive, MAST}
\software{\texttt{lightkurve} \citep{Lightkurve2018}, \texttt{transitleastsquares} \citep{Hippke2019}, \texttt{triceratops} \citep{Giacalone2021}, \texttt{celerite2} \citep{ForemanMackey2017}, \texttt{emcee} \citep{ForemanMackey2013}, \texttt{batman} \citep{Kreidberg2015}, \texttt{astropy} \citep{Astropy2013, Astropy2018, Astropy2022}}

\appendix

\section{Pipeline flowchart and classification details}
\label{app:flowchart}

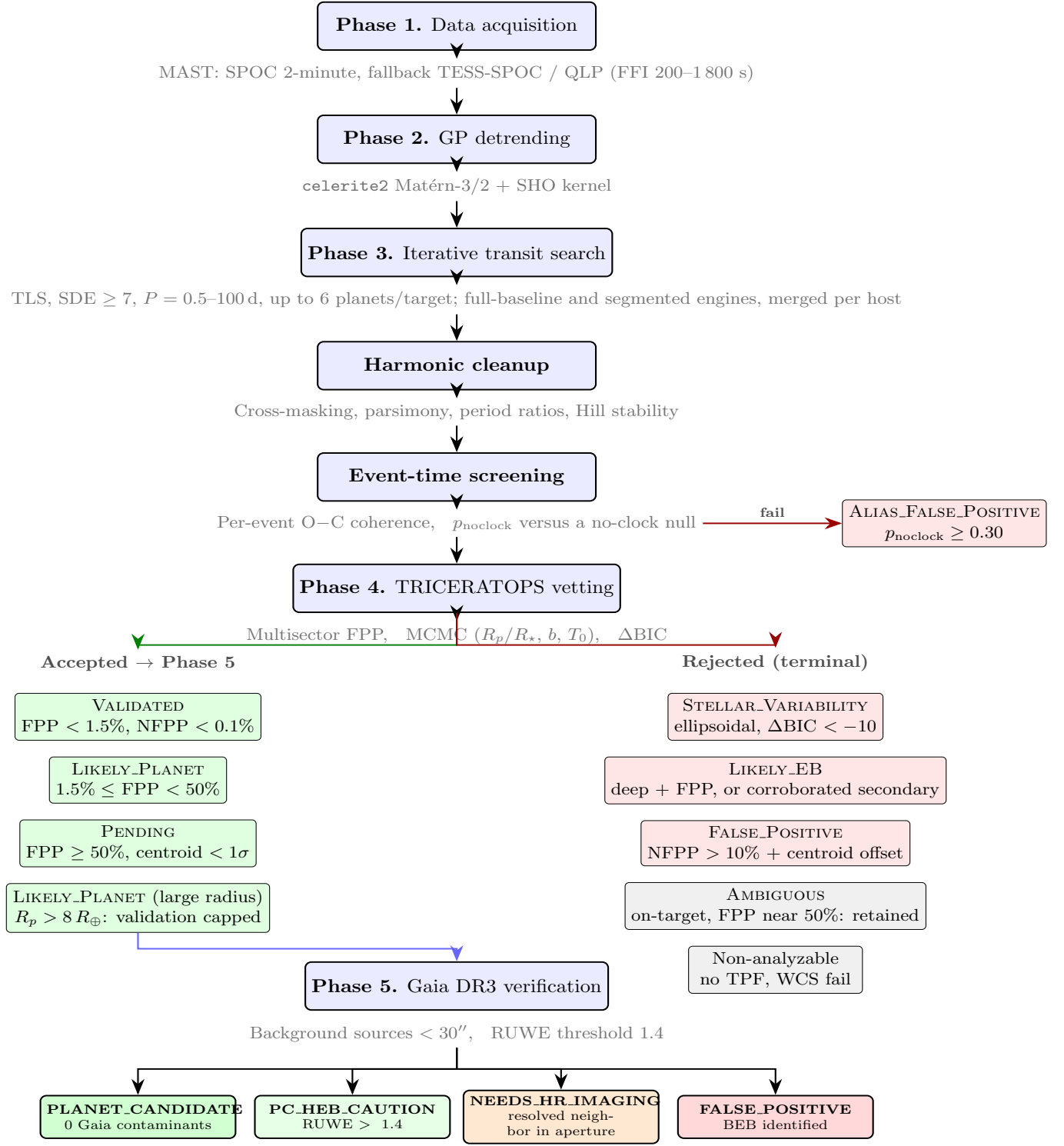
\begin{figure*}[!htbp]
\centering
\resizebox{\textwidth}{0.85\textheight}{%
\begin{tikzpicture}[
  node distance=0.55cm,
  phase/.style={draw, rounded corners=3pt, minimum height=0.9cm, minimum width=4.8cm,
                font=\normalsize, align=center, fill=blue!8, thick},
  method/.style={font=\small, text=black!55, align=center},
  accepted/.style={draw, rounded corners=2pt, minimum height=0.7cm, minimum width=3.0cm,
                   font=\small, align=center, fill=green!12},
  rejected/.style={draw, rounded corners=2pt, minimum height=0.7cm, minimum width=3.0cm,
                   font=\small, align=center, fill=red!10},
  final/.style={draw, rounded corners=3pt, minimum height=0.8cm, text width=3.15cm,
                font=\scriptsize, align=center, thick},
  lbl/.style={font=\small\bfseries, text=black!70},
  arr/.style={-{Stealth[length=3mm]}, thick},
]
\node[phase] (acq) {\textbf{Phase 1.} Data acquisition};
\node[method, below=0.15cm of acq] (acqm) {MAST: SPOC 2-minute, fallback TESS-SPOC / QLP (FFI 200--1\,800 s)};
\node[phase, below=0.5cm of acqm] (gp) {\textbf{Phase 2.} GP detrending};
\node[method, below=0.15cm of gp] (gpm) {\texttt{celerite2} Mat\'{e}rn-3/2 + SHO kernel};
\node[phase, below=0.5cm of gpm] (tls) {\textbf{Phase 3.} Iterative transit search};
\node[method, below=0.15cm of tls] (tlsm) {TLS, SDE $\geq 7$, $P = 0.5$--$100$\,d, up to 6 planets/target; full-baseline and segmented engines, merged per host};
\node[phase, below=0.5cm of tlsm] (hc) {\textbf{Harmonic cleanup}};
\node[method, below=0.15cm of hc] (hcm) {Cross-masking, parsimony, period ratios, Hill stability};
\node[phase, below=0.5cm of hcm] (etc) {\textbf{Event-time screening}};
\node[method, below=0.15cm of etc] (etcm) {Per-event O$-$C coherence, \; $p_{\rm noclock}$ versus a no-clock null};
\node[rejected, right=2.4cm of etcm, minimum width=3.4cm] (alias) {\textsc{Alias\_False\_Positive}\\$p_{\rm noclock} \geq 0.30$};
\node[phase, below=0.5cm of etcm] (tri) {\textbf{Phase 4.} TRICERATOPS vetting};
\node[method, below=0.15cm of tri] (trim) {Multisector FPP, \; MCMC ($R_p/R_\star$, $b$, $T_0$), \; $\Delta$BIC};
\draw[arr] (acq) -- (acqm); \draw[arr] (acqm) -- (gp); \draw[arr] (gp) -- (gpm);
\draw[arr] (gpm) -- (tls); \draw[arr] (tls) -- (tlsm); \draw[arr] (tlsm) -- (hc);
\draw[arr] (hc) -- (hcm); \draw[arr] (hcm) -- (etc); \draw[arr] (etc) -- (etcm); \draw[arr] (etcm) -- (tri); \draw[arr] (tri) -- (trim);
\draw[arr, red!60!black] (etcm.east) -- node[lbl, above, font=\scriptsize\bfseries] {\textsc{fail}} (alias.west);
\node[lbl, anchor=south] (albl) at ([shift={(-5.5cm,-1.2cm)}]tri.south) {Accepted $\rightarrow$ Phase~5};
\node[lbl, anchor=south] (rlbl) at ([shift={(5.5cm,-1.2cm)}]tri.south) {Rejected (terminal)};
\node[accepted, below=0.3cm of albl] (val) {\textsc{Validated}\\FPP $< 1.5\%$, NFPP $< 0.1\%$};
\node[accepted, below=0.3cm of val]  (lp)  {\textsc{Likely\_Planet}\\$1.5\% \leq$ FPP $< 50\%$};
\node[accepted, below=0.3cm of lp]   (ppv) {\textsc{Pending}\\FPP $\geq 50\%$, centroid $< 1\sigma$};
\node[accepted, below=0.3cm of ppv]  (lhj) {\textsc{Likely\_Planet} (large radius)\\$R_p > 8\,R_\oplus$: validation capped};
\node[rejected, below=0.3cm of rlbl] (sv)  {\textsc{Stellar\_Variability}\\ellipsoidal, $\Delta$BIC $< -10$};
\node[rejected, below=0.3cm of sv]   (leb) {\textsc{Likely\_EB}\\deep + FPP, or corroborated secondary};
\node[rejected, below=0.3cm of leb]  (fp4) {\textsc{False\_Positive}\\NFPP $> 10\%$ + centroid offset};
\node[draw, rounded corners=2pt, minimum height=0.6cm, minimum width=3.4cm,
      font=\small, align=center, fill=gray!12, below=0.3cm of fp4] (amb) {\textsc{Ambiguous}\\on-target, FPP near 50\%: retained};
\node[draw, rounded corners=2pt, minimum height=0.6cm, minimum width=3.0cm,
      font=\small, align=center, fill=gray!12, below=0.3cm of amb] (na) {Non-analyzable\\no TPF, WCS fail};
\draw[arr, green!50!black] (tri.south) -- ++(0,-0.6) -| (albl.north);
\draw[arr, red!60!black]   (tri.south) -- ++(0,-0.6) -| (rlbl.north);
\node[phase] (gaia) at ([shift={(0,-1.0cm)}]lhj.south -| tri.south) {\textbf{Phase 5.} Gaia DR3 verification};
\node[method, below=0.15cm of gaia] (gaiam) {Background sources $< 30\arcsec$, \; RUWE threshold 1.4};
\draw[arr, blue!60] (lhj.south) -- ++(0,-0.3) -| (gaia.north);
\node[final, fill=green!18] (pc)    at ([shift={(-5.5cm,-1.3cm)}]gaiam.south) {\textbf{PLANET\_CANDIDATE}\\0 Gaia contaminants};
\node[final, fill=green!10] (pcheb) at ([shift={(-1.8cm,-1.3cm)}]gaiam.south) {\textbf{PC\_HEB\_CAUTION}\\RUWE $> 1.4$};
\node[final, fill=orange!18](nhr)   at ([shift={(1.8cm,-1.3cm)}]gaiam.south)  {\textbf{NEEDS\_HR\_IMAGING}\\resolved neighbor in aperture};
\node[final, fill=red!15]  (fp5)    at ([shift={(5.5cm,-1.3cm)}]gaiam.south)  {\textbf{FALSE\_POSITIVE}\\BEB identified};
\draw[arr] (gaiam.south) -- ++(0,-0.4) -| (pc.north);
\draw[arr] (gaiam.south) -- ++(0,-0.4) -| (pcheb.north);
\draw[arr] (gaiam.south) -- ++(0,-0.4) -| (nhr.north);
\draw[arr] (gaiam.south) -- ++(0,-0.4) -| (fp5.north);
\end{tikzpicture}%
}%
\caption{Pipeline architecture from data acquisition to final classification. An event-time coherence screen removes window-function aliases ($p_{\rm noclock} \geq 0.30$, classified \textsc{Alias\_False\_Positive}) before TRICERATOPS vetting. Accepted vetting statuses (left, green) proceed to Gaia~DR3 verification; rejected statuses (right, red) are terminal; \textsc{Ambiguous} and non-analyzable signals (gray) are retained without promotion.}
\label{fig:pipeline_flowchart}
\end{figure*}

The classification stage of Sect.~\ref{sec:vetting} applies the following refinements relative to the default TRICERATOPS thresholds:

\begin{itemize}
    \item No impact parameter rejection: no candidate is rejected on the MCMC impact parameter $b$ alone, which the $b$--$R_p/R_\star$ degeneracy (Sect.~\ref{sec:limitations}) drives to $b \approx 1$ for shallow transits: three confirmed planets (GJ~3090\,b, \citealt{Almenara2022}; TOI-4529\,b, \citealt{Poultourtzidis2026}; TOI-206\,b, \citealt{Giacalone2022}) have pipeline MCMC $b \approx 1.09$ against literature $b = 0.38$, $0.32$, and $0.66$. No major survey uses a fixed $b$ threshold \citep{Thompson2018, Giacalone2021, Morton2012}.

    \item $R_p > 8\,R_\oplus$ bypass: for candidates with $R_p > 8\,R_\oplus$, where statistical validation is not possible (Sect.~\ref{sec:limitations}), the TRICERATOPS $\mathrm{prob}_{\mathrm{EB}}$ is not used in the classification decision and the candidate is capped at \texttt{LIKELY\_PLANET}.

    \item Centroid--NFPP consensus: a candidate is rejected on NFPP grounds only when the NFPP exceeds $10\%$ and the centroid offset itself exceeds $0.2$\,pixels. When the centroid offset is $< 1\sigma$ (signal on target) and all observational tests are clean (no secondary eclipse, odd-even $< 3\sigma$, $\mathrm{prob}_{\mathrm{EB}} < 30\%$), the candidate proceeds to background source analysis regardless of NFPP value. A candidate with $\mathrm{NFPP} > 10\%$ is a true false positive only 85\% of the time \citep[Sect.~5]{Giacalone2021}; the spatial centroid measurement complements the statistical prior.

    \item Stellar variability guard: on active stars where the BIC favors a constant model ($\Delta\mathrm{BIC} < -10$; \citealt{KassRaftery1995}), candidates are classified as \texttt{STELLAR\_VARIABILITY} instead of \texttt{LIKELY\_EB}.

    \item Depth threshold for eclipsing binaries: a signal is classified as \texttt{LIKELY\_EB} when its depth exceeds the fixed threshold of 50\,000\,ppm and the FPP is $\geq 50\%$ (Fig.~\ref{fig:pipeline_flowchart}); a depth above 50\,000\,ppm implies $R_p > 0.22\,R_\star$, larger than any planet on all but the smallest hosts. The depth cut is deliberately not used alone, because no fixed threshold is reliable across the M-dwarf radius range: 50\,000\,ppm is too aggressive on the smallest hosts, where a genuine $3\,R_\oplus$ transit on a $0.1\,R_\odot$ star already reaches ${\sim}75\,000$\,ppm, and too permissive on the largest, where a Jupiter-sized eclipsing companion on a $0.7\,R_\odot$ star produces only ${\sim}20\,000$\,ppm; the FPP condition supplies the discriminating evidence. The adaptive host-dependent ceiling of the search stage (Sect.~\ref{sec:detection}) complements this cut upstream.
\end{itemize}

For candidates with FPP $\geq 50\%$ (classified as \texttt{PENDING} by TRICERATOPS), Gaia governs the disposition: the absence of a resolvable contaminant yields \texttt{PLANET\_CANDIDATE} (the resolvable BEB scenario excluded independently of the FPP value). For candidates with FPP $< 50\%$ (\texttt{LIKELY\_PLANET}), Gaia can only upgrade, never downgrade \citep[consistent with the statistical validation already performed;][]{Giacalone2021}. Candidates with a resolved Gaia neighbor inside the photometric aperture, bright enough to host the signal, are flagged as \texttt{NEEDS\_HR\_IMAGING}, requiring high-resolution imaging or on-target photometry to localize the transit. This classification reflects aperture contamination, not signal quality: 24 confirmed planets in this sample also receive this flag (Sect.~\ref{sec:recovery_vetting}).

\section{Candidate ephemerides and full screening dispositions}
\label{app:candidate_tables}

Table~\ref{tab:ephemerides} lists the ephemerides and parameters of every recovery and new candidate; Table~\ref{tab:candidates_screening} gives the event-time screening dispositions of the tier~2 and tier~3 candidates and of the signals not retained as candidates. The recoveries and four new tier~1 candidates, the survey's strongest new signals, are summarized in the main text (Table~\ref{tab:screening_main}, Sect.~\ref{sec:candidates_revised}).

\begin{figure*}[!htbp]
\centering
\includegraphics[width=0.72\textwidth]{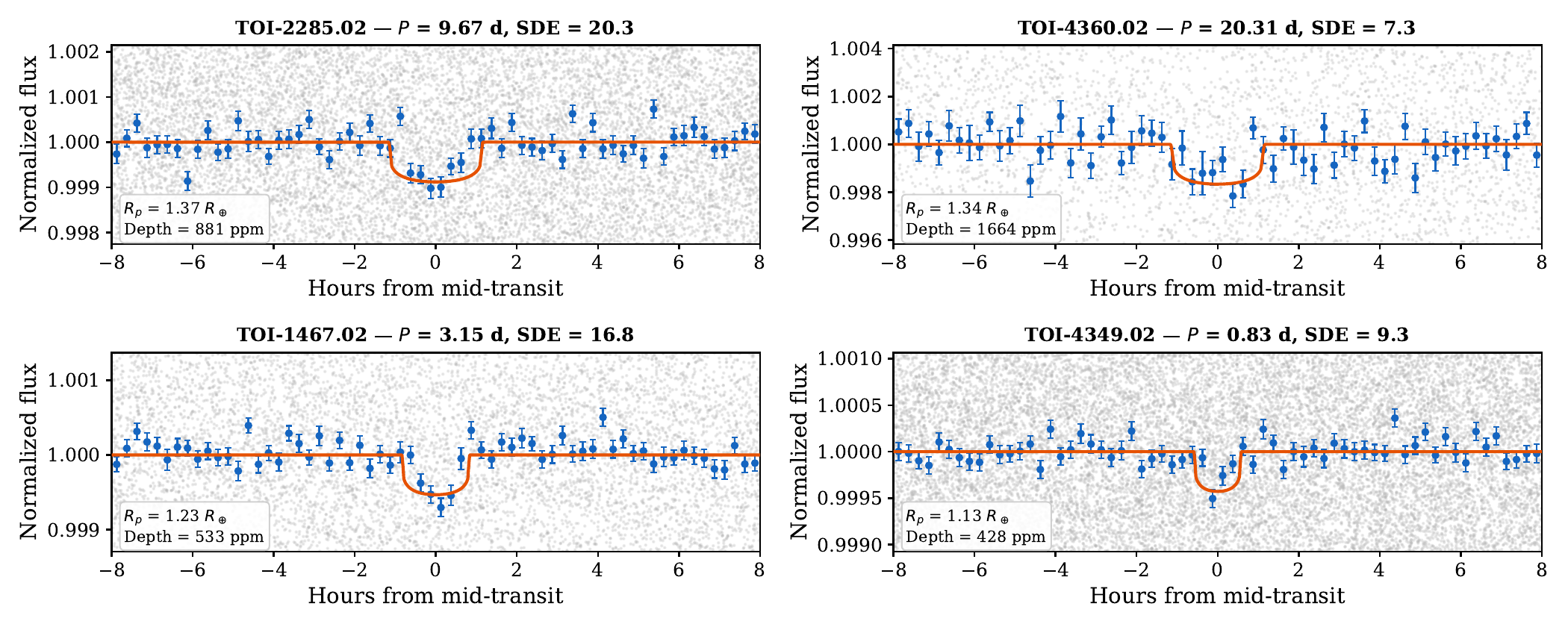}
\caption{Phase-folded TESS light curves of the two strongest tier~2 signals (left column; Sect.~\ref{sec:tier2}) and the two marginal tier~3 candidates (right column; Sect.~\ref{sec:tier3}); format as in Fig.~\ref{fig:grid_recoveries}. Gray: unbinned photometry; blue circles: binned data with $1\sigma$ error bars; orange: limb-darkened transit model. TOI-2285.02 is independently co-detected by \citet{Fukui2025} in the confirmed TOI-2285 system.}
\label{fig:grid_tier23}
\end{figure*}

\begin{table*}[!t]
\centering
\caption{Ephemerides and parameters of the cataloged-TOI recoveries and the new transit candidates, ordered as in the text (recoveries, then tier~1, tier~2, tier~3).}
\label{tab:ephemerides}
\scriptsize
\setlength{\tabcolsep}{3pt}
\begin{tabular}{@{}llccccccc l@{}}
\toprule
Designation & Host (TIC) & $P$ (d) & $T_0$ (BTJD) & Depth (ppm) & SDE & $R_p$ ($R_\oplus$) & $b$ & FPP / NFPP & Class \\
\midrule
\multicolumn{10}{@{}l}{\textit{Recoveries of cataloged TOIs (periods determined or recovered here) and the confirmed planet TOI-237\,c}}\\
TOI-2433.01 & 405763009 & $65.4590(3)$ & $1746.377(5)$ & $2\,308$ & $16.3$ & $1.82 \pm 0.21$ & $0.58$ & $31\% / 0$ & Rec. \\
TOI-4565.01 & 381897917 & $19.78578(1)$ & $1676.6153(8)$ & $1\,562$ & $22.6$ & $2.61^{+0.16}_{-0.17}$ & $0.56$ & $0.13\% / 0.12\%$ & Rec. \\
TOI-2293.01 & 71347873 & $6.068301(3)$ & $1847.8473(5)$ & $1\,225^{+92}_{-84}$ & $30.8$ & $1.91 \pm 0.12$ & $0.35$ & $6\% / 0.03\%$ & Rec. \\
TOI-4353.01 & 176797879 & $39.8991(3)$ & $1427.014(5)$ & $907$ & $14.9$ & $1.4 \pm 0.1$ & $0.14$ & $64\% / 0$ & Rec. \\
TOI-6492.01 & 47143972 & $33.2640(3)$ & $1530.330(5)$ & $36\,407$ & $14.3$ & $10.4$ & $0.50$ & $72\% / 28\%$ & Rec. \\
TOI-237\,c & 305048087 & $1.7448617(2)$ & $1355.5916(2)$ & $3\,352 \pm 200$ & $18.7$ & $1.21 \pm 0.04$ & $0.41$ & $40.0\% / 0\%$\textsuperscript{b} & Rec. (CP) \\
\multicolumn{10}{@{}l}{\textit{New candidates: tier 1}}\\
TOI-4556.02 & 107018378 & $2.4347553(4)$ & $1574.1102(3)$ & $2\,690 \pm 200$ & $36.0$ & $1.18^{+0.14}_{-0.13}$ & $0.52$ & $2.46\% / 2.40\%$ & 1 \\
TOI-6284.02 & 286763141 & $5.244499(2)$ & $1546.8709(6)$ & $298 \pm 90$ & $23.8$ & $0.89 \pm 0.13$ & $0.49$ & $6.7\% / 5.8\%$\textsuperscript{a} & 1 \\
TOI-6284.03 & 286763141 & $7.349312(2)$ & $1544.5112(5)$ & $465 \pm 125$ & $24.0$ & $1.04^{+0.16}_{-0.15}$ & $0.62$ & $24.9\% / 19.2\%$\textsuperscript{a} & 1 \\
TOI-4572.02 & 154940895 & $2.4495985(9)$ & $1817.0967(3)$ & $437 \pm 90$ & $18.5$ & $0.98^{+0.18}_{-0.17}$ & $0.61$ & $4.2\% / 1.9\%$ & 1 \\
\multicolumn{10}{@{}l}{\textit{New candidates: tier 2}}\\
TOI-2293.02 & 71347873 & $35.3560(3)$ & $1874.446(5)$ & $1\,207^{+286}_{-227}$ & $9.1$ & $1.89^{+0.23}_{-0.21}$ & $0.53$ & $50\% / 4.2\%$ & 2 \\
TOI-1467.02 & 240968774 & $3.154156(1)$ & $1767.4339(5)$ & $533 \pm 215$ & $16.8$ & $1.23 \pm 0.25$ & $1.10$ & $33\% / 7.2\%$ & 2 \\
TOI-2285.02 & 329148988 & $9.673244(5)$ & $1747.6299(6)$ & $881 \pm 270$ & $20.3$ & $1.37^{+0.25}_{-0.24}$ & $0.53$ & $29.8\% / 26.4\%$\textsuperscript{e} & 2 \\
TOI-4353.02 & 176797879 & $2.927759(1)$ & $1411.3407(6)$ & $284 \pm 90$ & $8.9$ & $0.73^{+0.16}_{-0.15}$ & $0.95$ & $6.2\% / 0.01\%$ & 2 \\
TOI-4508.02 & 170789802 & $2.2847558(5)$ & $1440.0223(4)$ & $823 \pm 200$ & $9.6$ & $1.01 \pm 0.22$ & $1.08$ & $38\% / 12\%$ & 2 \\
TOI-6578.02 & 144077896 & $1.2707381(3)$ & $1571.6952(2)$ & $900 \pm 230$ & $9.4$ & $1.75 \pm 0.36$ & $1.11$ & $12\% / 3.1\%$ & 2 \\
TOI-6578.03 & 144077896 & $3.456829(2)$ & $1572.2288(4)$ & $1\,282 \pm 350$ & $10.9$ & $1.97 \pm 0.35$ & $0.80$ & $24\% / 5.3\%$ & 2 \\
\multicolumn{10}{@{}l}{\textit{New candidates: tier 3}}\\
TOI-4360.02 & 52307802 & $20.30834(1)$ & $1327.122(1)$ & $1\,664 \pm 430$ & $7.3$ & $1.34 \pm 0.18$ & $0.46$ & $15.1\% / 10.9\%$\textsuperscript{c} & 3 \\
TOI-4349.02 & 313889049 & $0.8291551(2)$ & $1354.9202(3)$ & $428 \pm 110$ & $9.3$ & $1.13^{+0.25}_{-0.22}$ & $0.82$ & $31.6\% / 13.2\%$\textsuperscript{d} & 3 \\
\bottomrule
\end{tabular}
\par\smallskip
{\footnotesize\raggedright \textbf{Notes.} $^{\mathrm{a}}$TOI-6284 field cleared by Gemini speckle and a ground nearby-eclipsing-binary check, residual FPP the on-target EB. $^{\mathrm{b}}$TOI-237\,c is a confirmed planet, validated by \citet{Timmermans2026}; recovered blind here, its FPP a screen calibration point, not counted among the new candidates. $^{\mathrm{c}}$TOI-4360.02 SOAR speckle lowers the BEB probability $0.19\to0.02$, residual on-target EB. $^{\mathrm{d}}$TOI-4349.02 in-aperture neighbor excluded by Gemini speckle (tabulated NFPP is pre-imaging). $^{\mathrm{e}}$TOI-2285.02 is independently co-detected by \citet{Fukui2025}. Epochs $T_0$ in BTJD ($=$BJD$-2\,457\,000$). Parenthesized uncertainties on $P$/$T_0$ are $1\sigma$ on the last digit (weighted linear fit to the individual transit times); the four few-transit signals (TOI-2293.02, TOI-2433.01, TOI-4353.01, TOI-6492.01) carry baseline-limited values (15--40\,min mid-transit uncertainty at 2027.0). Depths are fixed-ephemeris fits on the SPOC 2-minute photometry; radii combine the fit posterior with the TIC radius \citep{Stassun2019} at a $5\%$ $\sigma_{R_\star}/R_\star$ floor. Recovery rows carry campaign detection-fit values, except TOI-2293.01 (fixed-ephemeris refit) and TOI-4353.01 (CHIRON-revised radius). The radius of TOI-6492.01, degenerate at its $3.6\%$ eclipse depth, is quoted without an uncertainty.\par}
\end{table*}

\begin{deluxetable*}{lcccccl}
\tabletypesize{\footnotesize}
\tablecaption{Event-time screening dispositions of the tier~2 and tier~3 candidates and of the signals not retained as candidates.\label{tab:candidates_screening}}
\tablehead{\colhead{Designation} & \colhead{$P$ (d)} & \colhead{SDE} & \colhead{$p_{\rm noclock}$} & \colhead{FPP} & \colhead{NFPP} & \colhead{Outcome}}
\startdata
TOI-2293.02 & 35.356  & 9.1  & FLAG\_LOWSENS  & $50\%$   & $4.2\%$  & tier~2 \\
TOI-1467.02 & 3.1542  & 16.8 & $0.005$ (PASS) & $33\%$   & $7.2\%$  & tier~2 \\
TOI-2285.02 & 9.6732  & 20.3 & $0.005$ (PASS) & $29.8\%$ & $26.4\%$ & tier~2 \\
TOI-4353.02 & 2.9277  & 8.9  & $0.235$ (FLAG) & $6.2\%$  & $0.01\%$ & tier~2 \\
TOI-4508.02 & 2.2848  & 9.6  & $0.055$ (FLAG) & $38\%$   & $12\%$   & tier~2 \\
TOI-6578.02 & 1.2707  & 9.4  & $0.008$ (PASS) & $12\%$   & $3.1\%$  & tier~2 \\
TOI-6578.03 & 3.4568  & 10.9 & $0.23$ (FLAG)  & $24\%$   & $5.3\%$  & tier~2 \\
TOI-4360.02 & 20.308  & 7.3  & $0.005$ (PASS) & $15.1\%$ & $10.9\%$ & tier~3 \\
TOI-4349.02 & 0.8292  & 9.3  & $0.20$ (FLAG)  & $31.6\%$ & $13.2\%$ & tier~3 \\
TOI-2495.02 & 3.247   & 10.7 & $<0.005$ (PASS) & \dots    & \dots    & Rejected \\
TOI-4559.02 & 9.277   & 9.2  & $0.035$ (PASS) & $100\%$  & $100\%$  & Rejected \\
TOI-4559.03 & 10.308  & 10.2 & $<0.005$ (PASS) & $100\%$  & $100\%$  & Rejected \\
TOI-663     & 11.892  & 5.6  & FLAG           & \dots    & \dots    & Indet. \\
TOI-4360 (54.34 d) & 54.34 & \dots & FLAG      & \dots    & \dots    & Demoted \\
TOI-4349    & 31.4934 & \dots  & PASS          & \dots    & \dots    & Demoted \\
\enddata
\tablecomments{Groups, top to bottom: seven tier~2 candidates (TOI-2285.02 independently co-detected by \citealt{Fukui2025}), two tier~3 candidates, three signals rejected on physical grounds despite a timing PASS, and three left indeterminate or demoted. Screen verdict $=$ timing-only $p_{\rm noclock}$ (PASS $\le 0.05$, FLAG $0.05$--$0.30$, FAIL $\ge 0.30$; FLAG\_LOWSENS for fewer than five timeable events; Sect.~\ref{sec:screening}), calibrated on TOI-521\,c (Sect.~\ref{sec:screening_calibration}). FPP and NFPP are the all-sector median screen values (Sect.~\ref{sec:vetting}); a dash marks a signal rejected or withdrawn before a probability was computed. Per-signal detail is in Sect.~\ref{sec:candidates_revised}.}
\end{deluxetable*}

\section{Full recovered confirmed planet catalog}
\label{app:recovered}

All 165 recovered confirmed planets, with the detected period $P_{\rm det}$ and full-baseline SDE, are provided as a machine-readable table in the electronic edition (Table~\ref{tab:app_recovered}); each $P_{\rm det}$ matches its ExoFOP catalog period to within $1\%$ (the recovery criterion, Sect.~\ref{sec:recovery_tls}), except TOI-5720.01, recovered at the $3P$ harmonic. The first rows, ordered by SDE, are shown below.

\begin{deluxetable}{lcrr}
\tabletypesize{\footnotesize}
\tablecaption{Recovered confirmed planets (excerpt; the full 165-entry catalog is available as a machine-readable table).\label{tab:app_recovered}}
\tablehead{\colhead{TOI} & \colhead{TIC} & \colhead{$P_{\rm det}$ (d)} & \colhead{SDE}}
\startdata
2119.01 & 236387002 & 7.2008  & 58.0 \\
6508.01 & 142277868 & 18.9928 & 54.7 \\
1752.01 & 287139872 & 0.9352  & 50.7 \\
5205.01 & 419411415 & 1.6307  & 43.5 \\
700.01  & 150428135 & 16.0512 & 42.1 \\
4666.01 & 165202476 & 2.9089  & 39.0 \\
$\vdots$ & \nodata & \nodata & \nodata \\
\enddata
\tablecomments{SDE is the full-baseline detection statistic. TOI-2015.01 and TOI-2267.01, at $\mathrm{SDE} = 5.7$, fall below the full-baseline threshold at the catalog period and enter the recovered set through the segmented search (Sect.~\ref{sec:segmented}). Full table (165 rows) in the electronic edition.}
\end{deluxetable}

\section{Full missed planet catalog}
\label{app:missed}

All 28 not-detected confirmed planets, with the raw SDE where a sub-threshold peak was found and the assigned cause, are provided as a machine-readable table (Table~\ref{tab:app_missed}); representative rows are shown below. Cause codes: the raw SDE of a sub-threshold peak; ``no signal at any period''; a dominant signal extracted first on the host at a different period (iterative masking); or the red-noise correction where a peak reaching $\mathrm{SDE_{raw}} \geq 7$ falls to $\mathrm{SDE_{eff}} = \mathrm{SDE_{raw}}/\sqrt{\beta_{\rm rn}} < 7$.

\begin{deluxetable*}{llccl}
\tabletypesize{\footnotesize}
\tablecaption{Not-detected confirmed planets (excerpt; the full 28-entry catalog is available as a machine-readable table).\label{tab:app_missed}}
\tablehead{\colhead{TOI} & \colhead{TIC} & \colhead{$P$ (d)} & \colhead{$N_{\rm sec}$} & \colhead{Cause}}
\startdata
1749.03 & 233602827 & 2.389  & 37 & dominant 4.49\,d signal \\
7393.01 & 165598669 & 3.337  & 8  & $\mathrm{SDE_{raw}} = 5.3 < 7$ \\
455.01  & 98796344  & 5.359  & 2  & triple star, aperture dilution \\
2221.01 & 441420236 & 8.454  & 3  & AU\,Mic, flare/spot modulation \\
1634.01 & 201186294 & 0.989  & 2  & $1.0$\,d Earth-rotation deadband \\
$\vdots$ & \nodata & \nodata & \nodata & \nodata \\
\enddata
\tablecomments{TOI-406.02 sits at a $2{:}1$ ratio to its brighter sibling TOI-406.01 and is credited to it (Sect.~\ref{sec:recovery}); planet~b of TOI-2285 (revised to $13.64$\,d by \citealt{Fukui2025}) is not recovered, the host's strongest peak being the tier-2 candidate TOI-2285.02 (Sect.~\ref{sec:tier2}). Full table (28 rows) in the electronic edition.}
\end{deluxetable*}

\section{End-to-end validation on known systems}
\label{app:validation}

\begin{deluxetable*}{llclll}
\tabletypesize{\footnotesize}
\tablecaption{End-to-end validation on ten known systems: one row per signal through the chain (homogeneous 2-minute SPOC photometry).\label{tab:validation}}
\tablehead{\colhead{Signal} & \colhead{$P$ (d)} & \colhead{Search} & \colhead{Screening (Sect.~\ref{sec:harmonic_cleanup})} & \colhead{Vetting (Sect.~\ref{sec:vetting})} & \colhead{Verification (Sect.~\ref{sec:verification})}}
\startdata
L\,98-59\,b            & 2.253  & both & \textsc{pass}          & \texttt{LIKELY\_PLANET} & \texttt{PC} \\
L\,98-59\,c            & 3.691  & both & \textsc{pass}          & \texttt{LIKELY\_PLANET} & \texttt{PC} \\
L\,98-59\,d            & 7.451  & segm.\ & \textsc{pass}          & \texttt{LIKELY\_PLANET} & \texttt{PC} \\
L\,98-59 ($2{:}1$ alias of c) & 7.381 & full, absorbed at merge & \nodata & \nodata & \nodata \\
L\,98-59 (off-target)  & 1.049  & segm.\ & \textsc{pass}          & \texttt{PENDING}        & \texttt{NEB\_AP.\ SUSPECT} \\
TOI-700\,b             & 9.977  & both & \textsc{pass}          & \texttt{LIKELY\_PLANET} & \texttt{PC} \\
TOI-700\,c             & 16.051 & both & \textsc{pass}          & \texttt{LIKELY\_PLANET} & \texttt{PC} \\
TOI-700\,e             & 27.810 & both & \textsc{pass}          & \texttt{PENDING} (degen.)\tablenotemark{a} & \texttt{NHR} \\
TOI-700\,d             & 37.424 & full & \textsc{pass}          & \texttt{LIKELY\_PLANET} & \texttt{PC} \\
GJ\,357\,b             & 3.931  & both & \textsc{pass}          & \texttt{LIKELY\_PLANET} & \texttt{PC} \\
GJ\,486\,b             & 1.467  & both & \textsc{pass}          & \texttt{LIKELY\_PLANET} & \texttt{PC} \\
Gliese\,12\,b          & 12.761 & full & \textsc{pass}          & \texttt{LIKELY\_PLANET} & \texttt{PC} \\
TOI-406.01             & 13.176 & both & \textsc{pass}          & \texttt{PENDING} (degen.)\tablenotemark{a} & \texttt{NHR} \\
TOI-782.01             & 8.024  & both & \textsc{pass}          & \texttt{LIKELY\_PLANET} & \texttt{PC} \\
TOI-6086.01            & 1.389  & both & \textsc{flag}          & \texttt{LIKELY\_PLANET} & \texttt{PC} \\
GJ\,3473\,b            & 1.198  & both & \textsc{pass}          & \texttt{LIKELY\_PLANET} & \texttt{PC} \\
GJ\,3473 (spurious)    & 0.529  & full & \textbf{removed, parsimony cap} & \nodata & \nodata \\
GJ\,3473 (window alias)& 0.628  & full & \textbf{\textsc{fail}} ($p_{\rm noclock}=0.52$) & \texttt{ALIAS\_FALSE\_POS.} & \nodata \\
LP\,791-18\,b          & 0.948  & both & \textsc{pass}          & \texttt{PENDING}        & \texttt{NHR} \\
LP\,791-18\,c          & 4.990  & both & \textsc{pass}          & \texttt{LIKELY\_PLANET} & \texttt{PC} \\
LP\,791-18 (spurious)  & 34.343 & full & \textsc{flag\_lowsens} & \textbf{\texttt{FALSE\_POSITIVE}} & \nodata \\
LP\,791-18 (spurious)  & 39.799 & full & \textsc{flag\_lowsens} & \textbf{\texttt{FALSE\_POSITIVE}} & \nodata \\
LP\,791-18 (spurious)  & 64.234 & full & \textsc{flag\_lowsens} & \textbf{\texttt{FALSE\_POSITIVE}} & \nodata \\
\textbf{16 genuine}    & \nodata & \textbf{13 both, 2 full, 1 segm.} & \textbf{15 \textsc{pass}, 1 \textsc{flag}} & \textbf{13 \texttt{LP}, 3 \texttt{PENDING}} & \textbf{13 \texttt{PC}, 3 \texttt{NHR}} \\
\textbf{7 spurious}    & \nodata & \nodata & \multicolumn{3}{l}{\textbf{all removed: 1 merge, 1 cleanup, 1 event-time, 3 vetting, 1 verification}} \\
\enddata
\tablenotetext{a}{The degenerate joint folds (TOI-700\,e and TOI-406.01, \citealt{Gilbert2023}) and the off-target L\,98-59 $1.049$\,d signal ($\rho = +0.67$, NFPP $41\%$) are detailed in Sect.~\ref{sec:validation_known}.}
\tablecomments{One row per signal through the chain; ``\nodata'' marks stages not reached after a terminal removal (bold). ``Search'': both/full/segm.\ $=$ recovered by both engines, the full-baseline search alone, or the segmented search alone (Sect.~\ref{sec:segmented}).}
\end{deluxetable*}

\section{Vetting checks and follow-up target list}
\label{app:pc_phasefolds}

Three companion catalogs collect the ephemerides, observability (J2000 RA/Dec, $T_{\rm mag}$), and the full vetting-check set (transit duration and depth, SDE, FPP/NFPP, $p_{\rm noclock}$, odd-even difference, and difference-image centroid offset) as a reference for ground-based follow-up: Table~\ref{tab:checks_recoveries} for the five recovered known TOIs whose periods are determined or recovered here (Sect.~\ref{sec:recoveries_period}), Table~\ref{tab:checks_candidates} for the 13 new transit candidates in three confidence tiers (Sect.~\ref{sec:candidates_revised}), and Table~\ref{tab:checks_pc} for the 87 recovered planet candidates (Sect.~\ref{sec:followup}), ordered host-by-host. The full tables are provided in machine-readable form in the electronic edition; excerpts with a reduced column set are shown below. The background eclipsing binary scenarios of the recovered candidates are excluded by Gaia~DR3 (Sect.~\ref{sec:followup}).

\begin{deluxetable*}{llccl}
\tabletypesize{\footnotesize}
\tablecaption{Recovered known TOIs with periods determined or recovered here: follow-up reference (excerpt; full machine-readable table in the electronic edition).\label{tab:checks_recoveries}}
\tablehead{\colhead{TOI} & \colhead{TIC} & \colhead{$P$ (d)} & \colhead{$R_p$ ($R_\oplus$)} & \colhead{Next step}}
\startdata
TOI-2433.01 & 405763009 & 65.459 & 1.82 & RV; clean field; TESS S119--121 \\
TOI-4565.01 & 381897917 & 19.786 & 2.61 & Ground transit; TESS S104 \\
TOI-2293.01 & 71347873  & 6.068  & 1.91 & RV mass; EB excluded (TRES) \\
TOI-4353.01 & 176797879 & 39.899 & 1.40 & RV or transit shape; near-Earth, habitable zone \\
TOI-6492.01 & 47143972  & 33.264 & 10.4 & RV; deep degenerate eclipse \\
\enddata
\end{deluxetable*}

\begin{deluxetable*}{llccl}
\tabletypesize{\footnotesize}
\tablecaption{The 13 new transit candidates in three confidence tiers, with the co-recovered confirmed planet TOI-237\,c treated separately: follow-up reference (excerpt; full machine-readable table in the electronic edition).\label{tab:checks_candidates}}
\tablehead{\colhead{TOI} & \colhead{TIC} & \colhead{$P$ (d)} & \colhead{Tier} & \colhead{Next step}}
\startdata
TOI-4556.02 & 107018378 & 2.435 & 1 & Imaging / ground transit; TESS S114 \\
TOI-6284.02 & 286763141 & 5.244 & 1 & RV or transit shape (on-target EB) \\
TOI-6284.03 & 286763141 & 7.349 & 1 & Field cleared; ground NEB $+$ speckle \\
TOI-4572.02 & 154940895 & 2.450 & 1 & Imaging; EB excluded (TRES) \\
TOI-2285.02 & 329148988 & 9.673 & 2 & HR imaging; co-detected (Fukui 2025) \\
TOI-4360.02 & 52307802  & 20.308 & 3 & Extended baseline $+$ HR \\
$\vdots$ & \nodata & \nodata & \nodata & \nodata \\
\enddata
\end{deluxetable*}

\begin{deluxetable*}{llcccl}
\tabletypesize{\footnotesize}
\tablecaption{Recovered planet candidates: follow-up reference (excerpt of 87; full machine-readable table in the electronic edition).\label{tab:checks_pc}}
\tablehead{\colhead{TOI} & \colhead{TIC} & \colhead{$P$ (d)} & \colhead{Depth (ppm)} & \colhead{$R_p$ ($R_\oplus$)} & \colhead{Disp.}}
\startdata
TOI-5983.01 & 356871098 & 1.027  & 696  & 1.64 & PC \\
TOI-789.03  & 300710077 & 8.043  & 784  & 1.39 & PC \\
TOI-3288.01 & 79920467  & 1.434  & 30\,460 & 10.83 & PC \\
TOI-873.01  & 237920046 & 5.931  & 979  & 1.78 & PC \\
$\vdots$ & \nodata & \nodata & \nodata & \nodata & \nodata \\
\enddata
\tablecomments{All quantities are from the homogeneous 2-minute reduction. Full 87-entry table, with RA/Dec, $T_{\rm mag}$, $T_0$, $T_{14}$, SDE, FPP/NFPP, $p_{\rm noclock}$, odd-even, and centroid columns, in the electronic edition.}
\end{deluxetable*}

\clearpage
\bibliographystyle{aasjournal}
\bibliography{references}

\end{document}